\documentclass[aps,pra,twocolumn,superscriptaddress]{revtex4-2}
\usepackage{amsmath,amssymb,amsfonts,bbm,graphicx,times,psfrag}
\usepackage[pdftex]{color}
\usepackage{amsmath,bm}
\usepackage[colorlinks, linkcolor=blue, citecolor=red, urlcolor=red, breaklinks=true]{hyperref}
\usepackage{microtype}

\usepackage{amsthm}
\usepackage{csquotes}
\usepackage{amsthm}\newcommand{\be} {\begin{equation}}
\newcommand{\ee} {\end{equation}}

\usepackage{color}
\usepackage{mathrsfs}
\newcommand{\bla}{bla\\bla\\bla\\bla\\bla}

\begin{document}
\title{Entropy production rate and correlations of cavity magnomechanical system}
\author{Collins O. Edet}
\affiliation{Institute of Engineering Mathematics, Universiti Malaysia Perlis, 02600 Arau, Perlis, Malaysia}
\affiliation{Department of Physics, University of Cross River State, Calabar, Nigeria}
\author{Muhammad Asjad}\email{asjad\_qau@yahoo.com}
\affiliation{Department of Mathematics, Khalifa University, Abu Dhabi 127788, United Arab Emirates}
\author{Denys Dutykh}
\affiliation{Department of Mathematics, Khalifa University, Abu Dhabi 127788, United Arab Emirates}
\affiliation{Causal Dynamics Pty Ltd, Perth, Australia}
\author{Norshamsuri Ali}
\affiliation{Advanced Communication Engineering (ACE) Centre of Excellence, Universiti Malaysia Perlis, 01000 Kangar, Perlis, Malaysia}
\author{Obinna Abah}\email{obinna.abah@newcastle.ac.uk}
\affiliation{School of Mathematics, Statistics and Physics, Newcastle University, Newcastle upon Tyne NE1 7RU, United Kingdom}
\begin{abstract}
We present the irreversibility generated by a stationary cavity magnomechanical system composed of a yttrium iron garnet (YIG) sphere with a diameter of a few hundred micrometers inside a microwave cavity. In this system, the magnons, i.e., collective spin excitations in the sphere, are coupled to the cavity photon mode via magnetic dipole interaction and to the phonon mode via magnetostrictive force (optomechanical-like). We employ the quantum phase space formulation of the entropy change to evaluate the steady-state entropy production rate and associated quantum correlation in the system. We find that the behavior of the entropy flow between the cavity photon mode and the phonon mode is determined by the magnon-photon coupling and the cavity photon dissipation rate. Interestingly, the entropy production rate can increase/decrease depending on the strength of the magnon-photon coupling and the detuning parameters. We further show that the amount of correlations between the magnon and phonon modes is linked to the irreversibility generated in the system for small magnon-photon coupling. Our results demonstrate the possibility of exploring irreversibility in driven magnon-based hybrid quantum systems and open a promising route for quantum thermal applications.
\end{abstract}
\maketitle

\section{Introduction}
Hybrid quantum systems play a crucial role in the advancement of quantum technologies \citep{xiang2013hybrid,bauer2016hybrid,lihybrid:2017,lachance2019hybrid,clerk2020hybrid, callison2022hybrid}. In the last decade, remarkable progress has been made in this field with applications spanning through quantum computing~\cite{clerk2020hybrid,potter2022entanglement}, quantum simulation~\cite{wallquist2009hybrid}, quantum communication~\cite{callison2022hybrid}, quantum sensing~\cite{Degen2017}, and quantum thermodynamics \cite{myers2022quantum}. 
A typical example  is the cavity optomechanical system, which combines mechanical degrees of freedom with electromagnetic (EM) cavities~\citep{aspelmeyerbook:2014}. 
 Recently, hybrid quantum systems based on magnons, quanta of collective spin excitations in ordered ferrimagnetic materials, e.g. yttrium iron garnet (YIG), have attracted considerable attention due to great frequency tunability and very good coherence~ \cite{lachance2019hybrid}. Magnons can be coherently coupled to different degrees of freedom, such as phonons via the magnetostrictive force~\cite{zhang2016cavity}, microwave photons via the magnetic dipole interaction~\cite{lachance2019hybrid,Tabuchi2014,Bai2015}, optical photons~\cite{Osada2018,Haigh2016} and  superconducting qubits~\cite{tabuchi2015coherent,Wolski2020,Li2022}. The realization of this cavity magnomechanical system, a system of photon-magnon-phonon interaction with YIG spheres interacting with microwave cavities has open up possible applications in the preparation of macroscopic quantum states~\cite{Sun2021}, generation of squeezed states~\cite{Huang2023}, the generation of high-performance detectors \cite{potts2021dynamical},  entanglement generation \citep{li2018magnon,Liu2023}, ground state cooling \cite{Asjad2022}, and quantum information processing~\cite{Tan2019,Li2020,Li2021,Yuan2022}.

Quantum technological and nanofabrication advancements have motivated the design of microscopic and coherent thermodynamic machines -- quantum thermal machines~\cite{myers2022quantum}, as well as investigating the interplay between quantum information and thermodynamic processes~\cite{Goold2016}. Optomechanical thermal machines have been proposed in different configurations~\cite{zhang2014quantum,zhang2014theory,Zhang2017,Dechant2015,Serafini2020,Bathaee2016,Naseem2019}. The irreversibility (such as, friction, disorder) that influences the machine thermodynamic processes/performance can be quantified by entropy production \citep{tolman1948irreversible,sethna2021statistical}. Thus, quantifying the degree of irreversible entropy generated in a dynamic process is useful for the distinctive description of the  non-equilibrium processes, and decreasing it, enhances a thermal machine efficiency \citep{landi2021irreversible}. Based on quantum phase-space method, the measure of quantifying the irreversible entropy production of quantum systems that interact with nonequilibrium reservoirs has been formulated in \cite{brunelli2016irreversibility, santos2017wigner, santos2018spin} and experimentally verified in two distinct setup - an optomechanical system and a driven Bose-Einstein condensate coupled to a high finesse cavity \cite{brunelli2018experimental}. The effect of self-correlation on irreversible entropy production rate in a parametrically driven dissipative system has been investigated \cite{shahidani2022irreversible}. It has been recently shown that the presence of nonlinearity via an optical parametric oscillator placed inside the cavity optomechanical system influences the  stationary state entropy production rate~\cite{abah2023}. Moreover, it has been established that the entropy produced in a bipartite quantum system is related to the amount of correlations shared by its subsystems~\cite{brunelli2016irreversibility}. 

Here, we investigate the generation of irreversibility in a hybrid cavity magnomechanical setup comprising a microwave cavity and a YIG sphere. In this system, magnons are simultaneously coupled to the phonons of the vibrational sphere via magnetostrictive interaction and to the cavity photons via magnetic-dipole interaction, while there is no direct interaction between the cavity mode and the mechanical mode. We find that the magnon-photon coupling and cavity photon dissipation rate influence the entropy production rate. Furthermore, we demonstrate that the amount of correlations in the cavity magnomechanical system deviates from the steady-state entropy production rate for large magnon-photon coupling.

The rest of this paper is organized as follows. In Section \ref{model}, we present the physical model of the cavity magnomechanical system. We derive the linearized Hamiltonian of the system via quantum Langevin equations of motion and standard linearization techniques. In Section \ref{entropy_prod_rate}, the stationary entropy production rate and quantum correlation quantified by mutual information are presented using the experimentally feasible parameters. 
Finally, the conclusions are summarized in Section \ref{conclusion}.

\section{Cavity Magnomechanical  model}\label{model}
\begin{figure}[t!]
\includegraphics[width=.9 \columnwidth]{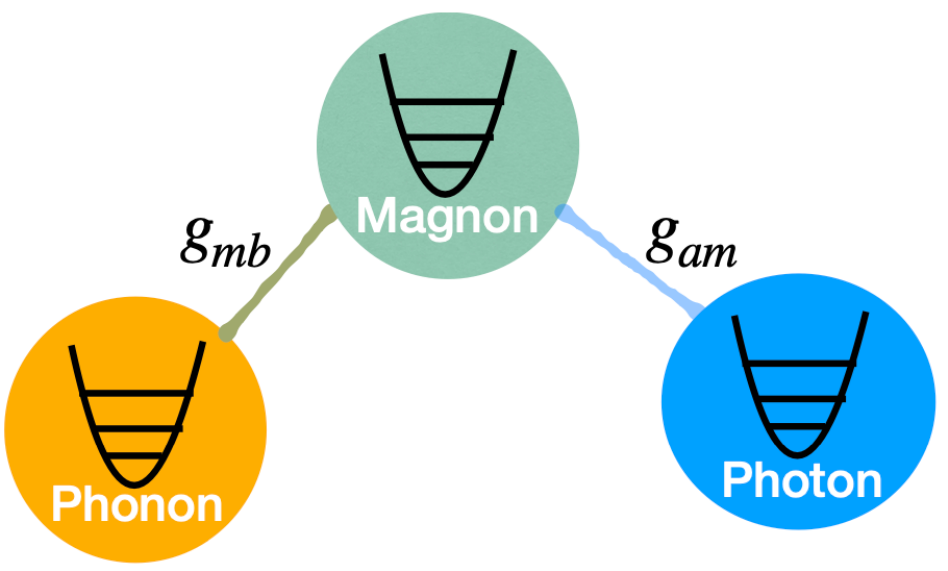} 
\caption{Diagrammatic  representation of a  cavity magnomechanical system  consisting of photon, magnon, and phonon modes. The magnon-photon interaction strength and the magnon-phonon interaction of coupling strength are denoted by $g_{\rm am}$ and $g_{\rm mb}$ respectively.}      
\label{fig1}
\end{figure}
We consider a hybrid cavity magnomechanical system, which consists of a microwave cavity and a small sphere (a one  $mm$-diameter, highly-polished YIG sphere is considered in Ref. \cite{wang2016magnon, ding2020ground}). The YIG sphere is positioned close to the maximal microwave magnetic field of the cavity mode, and a variable external magnetic field $H$ in the $z$-axis is added to establish the magnon–photon coupling~\cite{wang2016magnon, wang2018bistability}. The coupling rate can be tailored by adjusting the position of the sphere.
The magnons couple to phonons via the magnetostrictive effect. The vibrational modes (phonons) result from the geometric deformation of the YIG sphere because the magnon excitation inside the YIG sphere induces a varying magnetization. The magnomechanical coupling can be enhanced by directly driving the magnon mode with a microwave source \citep{zhang2016cavity}. In addition, the size of the sphere considered is much smaller than the wavelength of the microwave field, such that the interaction between cavity microwave photons and phonons can be neglected (i.e. the radiation pressure effect is negligible).
Thus, the system has three modes: cavity photon, magnon, and phonon modes, which can be schematically depicted by the equivalent coupled harmonic oscillator model as shown in Fig.~\ref{fig1}.
 
 The Hamiltonian of the hybrid quantum system under rotating-wave approximation in a frame rotating with the frequency $\omega_{\rm d}$ of the driving field  can be expressed as ($\hbar\!=\!1$) \cite{li2018magnon, ding2020ground}:
\begin{eqnarray}\label{freehamil}
\hat{H}&:=&\Delta_{\rm a}\hat{a}^{\dagger}\hat{a}+ \Omega_{\rm b}\hat{b}^{\dagger}\hat{b} +\Delta_{\rm m}\hat{m}^{\dagger}\hat{m}+g_{\rm  am} (\hat{a} \hat{m}^{\dagger}+  \hat{a}^{\dagger}\hat{m}) \nonumber \\
&+& g_{\rm mb}\, \hat{m}^{\dagger} \hat{m} (\hat{b} + \hat{b}^{\dagger}) +{\rm i}  (\Omega_{\rm d}\hat{m}^{\dagger}- \Omega_{\rm d}^{\ast}\hat{m}),
\end{eqnarray}
where the bosonic annihilation (creation) operators $\hat{a}\, (\hat{a}^\dagger)$, $\hat{b}\, (\hat{b}^\dagger)$ and $\hat{m}\, (\hat{m}^\dagger)$ denote, respectively, cavity photon, phonon and maganon modes whose resonant frequencies are taken to be $\omega_{\rm a}$, $\Omega_{\rm b}$, and $\omega_{\rm m}$ corrospondingly. The  detuning parameters $\Delta_{\rm a}:\!=\!\omega_{\rm a}-\omega_{\rm d}$, where $\omega_{\rm a}/2\pi\!=\! 10$ GHz \cite{zhang2016cavity}, and  $\Delta_{\rm m}:\!=\!\omega_{\rm m} -\omega_{\rm d}$. The uniform magnon mode frequency in the YIG sphere is $\omega_{\rm m}\!=\!\gamma_g H$, where $\gamma_g/2\pi\!=\!28$ GHz/T is the gyromagnetic ratio, and we set $\omega_{\rm m}$ at the Kittel mode frequency \cite{kittel1948theory}, which can lead to cavity polaritons by strongly coupling magnon and cavity photons. The parameters $g_{\rm am}$ and $g_{\rm mb}$ are the optomagnon  (photon-magnon) and magnomechanical (magnon-phonon) interaction coupling strengths respectively. The last term, $(\Omega_{d}\hat{m}^{\dagger}- \Omega_{d}^{\ast}\hat{m})$ in the Hamiltonian describes the external driving of the magnon mode. The Rabi frequency $\Omega_{d}\! \equiv \!\frac{\sqrt{5}}{4} \gamma_{g} \sqrt{N_t} B_0$ (assuming  low-lying excitations) denotes  the  coupling strength of the drive magnetic field \cite{li2018magnon}. The amplitude
and frequency of the drive field are $B_0$ and $\omega_{\rm d}$, respectively, and the total number of spins $N_t\!=\!\rho\, V$, where $V$ is the volume of the  sphere and $\rho\!=\!4.22 \times 10^{27}$m$^{-3}$ is the spin density of the YIG. Each of the modes is coupled to an independent noise reservoir, their energy decay rates are $\gamma_{a}$, $\gamma_{\rm m}$, and $\gamma_{\rm b}$, for photon, magnon and phonon respectively.
Experimentally, $g_{\rm mb}$ is extremely weak \cite{zhang2016cavity}, but the magnomechanical interaction can be enhanced by driving the magnon mode with a strong microwave field  \cite{wang2016magnon, wang2018bistability}.  The magnon-photon coupling rate $g_{\rm am}$ can be larger than the dissipation rates of the cavity and magnon modes, $\gamma_{\rm a}$  and $\gamma_{\rm m}$, entering into the strong coupling regime, $g_{\rm am}> \max\{\gamma_{\rm a},\gamma_{\rm m}\}$ \cite{Tabuchi2014,Bai2015}. 

As a result of strong driving, the Hamiltonian in Eq.~(\ref{freehamil}) can be linearized around the  coherent steady-state amplitude: $\hat{\mathscr{O}} \rightarrow  \mathscr{O}_{\rm s} + \hat{\mathscr{O}}$ ($\mathscr{O}$ \! $\in$ \! \{$a$, $b$, $m$\}), where  $\mathscr{O}_{\rm s}$ and the operators $\hat{\mathscr{O}}$, represent the steady-state amplitudes and quantum fluctuations of the corresponding mode. 
We have the steady-state amplitudes
\begin{equation}
     m_{\rm s} = \frac{\Omega_{\rm d} ({\rm i} \Delta_{\rm a}+\gamma_{a})}{g_{\rm am}^{2} + ({\rm i} \Delta_{\rm m}+\gamma_{\rm m}) ({\rm i} \Delta_{\rm a}+\gamma_{a})},
\end{equation}
and $ b_{\rm s} = - {\rm i}\, g_{\rm mb}	\left|  m_{\rm s} \right|^2/({\rm i}\,  \Omega_{\rm b} +\gamma_{\rm b})$.
Then, the linearized Hamiltonian can be derived as
\begin{eqnarray}\label{Hl}
\hat{H}_\text{lin}&=& \Delta_{\rm a} \hat{a}^{\dagger} \hat{a}+ \Omega_{\rm b} \hat{b}^{\dagger} \hat{b}+ \Tilde{\Delta}_{\rm m} \hat{m}^{\dagger} \hat{m} +g_{\rm am} (\hat{a} \hat{m}^{\dagger}+ \hat{a}^{\dagger} \hat{m})\nonumber\\
&+&(\mathscr{G}_{\rm mb}^{\ast} \hat{m}+\mathscr{G}_{\rm mb}\,  \hat{m}^{\dagger})(\hat{b}+\hat{b}^{\dagger}), 
\end{eqnarray}
where the enhanced magnon-phonon coupling $\mathscr{G}_{\rm mb}\! =\!g_{\rm mb} m_{\rm s}$, and $\Tilde{\Delta}_{\rm m}=\Delta_{\rm m}- g_{\rm mb} (b_{\rm s} + b^*_{\rm s})$ is the effective magnon detuning incorporating the magnetostriction.  For the considered parameters, $ g_{\rm mb} (b_{\rm s}  + b^*_{\rm s}  )\ll \Delta_{\rm m}$, so we can have, $\Tilde{\Delta}_{\rm m} \approx \Delta_{\rm m}$. 
Since the driving field affects $m_{\rm s}$, we can improve $\mathscr{G}_{\rm mb}$ by adjusting the external driving field $\Omega_d$.  

From the Hamiltonian in Eq. (\ref{Hl}), we obtain the quantum Langevin equations as;
\begin{eqnarray} \label{QLE}
\dot{\hat{a}}&=& -\left({\rm i} \Delta_{\rm a}+\gamma_{a} \right) \hat{a}- {\rm i} g_{\rm am} \hat{m}-\sqrt{2\gamma_{a}} \hat{a}_{\rm in},\nonumber \\
\dot{\hat{m}}&=&-\left({\rm i} \Tilde{\Delta}_{\rm m}+\gamma_{\rm m} \right) \hat{m}- {\rm i} g_{\rm am} \hat{a}-{\rm i} \mathscr{G}_{\rm mb}(\hat{b}+\hat{b}^{\dagger})-\sqrt{2\gamma_{\rm m}} \hat{m}_{\rm in},\nonumber \\ 
\dot{\hat{b}}&=&-\left({\rm i} \Omega_{\rm b} +\gamma_{\rm b}\right) \hat{b}-{\rm i} (\mathscr{G}_{\rm mb} \hat{m}^{\dagger}+\mathscr{G}_{\rm mb}^{*} \hat{m})-\sqrt{2\gamma_{\rm b}} \hat{b}_{\rm in}, 
\end{eqnarray}
where $\hat{f}_{\rm in}\! \in \{\!\hat{a}_{\rm in}, \hat{m}_{\rm in} ,\hat{b}_{\rm in}\}$ are  input noise operators for the cavity, magnon and mechanical modes, respectively, which are zero mean and characterized by the following correlation functions \cite{Gardiner2004}: $\langle \hat{f}_{\rm in} (t)  \hat{f}_{\rm in}^{\dagger} (t^{\prime}) \rangle=(\mathcal{N}_{k}+1)\delta(t-t^{\prime}) \, \text{and} \, \langle \hat{f}_{\rm in}^{\dagger} (t)  \hat{f}_{\rm in} (t^{\prime} )\rangle=\mathcal{N}_{k} \delta(t-t^{\prime})$, where $\mathcal{N}_{k}=1/\left({\rm e}^{\hbar \omega_{k}/k_\text{B}T}-1\right)$ $\left(k\!\in \! \{\rm a, m \}\right)$, are the equilibrium mean thermal photon,  and magnon number, respectively, while $\mathcal{N}_{\rm b}\!=\!1/\left({\rm e}^{\hbar\Omega_b/k_\text{B}T} - 1\right)$ is the equilibrium mean thermal phonon number, $T$ is the environmental temperature  and $k_\text{B}$ is the Boltzmann constant .
Eq.~(\ref{QLE}) represents the evolution of the fluctuation incorporating the interplay with the environment via the noise operators \cite{Asjad2022}. The photon number and magnon occupation number are approximately zero, i.e., $\mathcal{N}_{\rm a, m}\approx 0,$ due to the high frequencies of their modes. 

Since the nature of the noise is Gaussian, all the information is contained in the first and second-order moments of the operators.
In particular, it is convenient to introduce the quadratures $\hat{x}$ and $\hat{y}$ 
 of the photon, magnon, and phonon  modes by using the relation $\hat{\mathscr{O}}= (\hat{x}_\mathscr{O}+{\rm i} \hat{y}_\mathscr{O})/\sqrt{2}$ with $\mathscr{O} \in \{ \rm a,m, b\}$, and elements of the corresponding covariance matrix are defined as $\mathscr{V}_{ij} =\frac{1}{2}\langle \mathcal{R}_{i} (\infty) \mathcal{R}_{j} (\infty) +\mathcal{R}_{j} (\infty) \mathcal{R}_{i} (\infty)\rangle$, ($i, j\! \in \!\{ 1,2,3,4,5,6\}$).
Here, $\mathcal{R}_{i} (t): =[\hat{x}_{\rm a},\hat{y}_{\rm a}, \hat{x}_{\rm m}, \hat{y}_{\rm m},\hat{x}_{\rm b}, \hat{y}_{\rm b}]^\top$ is the column vector of quadratures and the stationary covariance matrix is obtained by solving the algebraic equation $\mathscr{A}^\top   \mathscr{V}  +  \mathscr{A}  \mathscr{V} + \mathscr{D}=0$, where 
 $\mathscr{D}\!=\!\text{diag}\left(\gamma_a, \gamma_{\rm a}, \gamma_{\rm m},\gamma_{\rm m}, \gamma_{\rm b} (2\mathcal{N}_{\rm b} +1),\gamma_{\rm b} (2\mathcal{N}_{\rm b} +1) \right)$ and the drift matrix $\mathscr{A}$ is expressed as:
 \begin{equation}\label{driftmatrix}
\mathscr{A}= \begin{pmatrix}
        -\gamma_{\rm a}& \Delta_{\rm a}& 0&g_{\rm am}&0&0\\
        -\Delta_{\rm a}& -\gamma_{\rm a}&-g_{\rm am}&0&0&0\\
         0& g_{\rm am}&-\gamma_{\rm m}&\Tilde{\Delta}_{\rm m}&-\mathscr{G}_{\rm mb}&0\\
          -g_{\rm am}& 0 &-\Tilde{\Delta}_{\rm m}&-\gamma_{\rm m}&0&0\\
           0& 0 & 0 & 0 & 0 & \Omega_{\rm b} \\
         0 & 0 &0 &\mathscr{G}_{\rm mb}&-\Omega_{\rm b} & -\gamma_{\rm b}
\end{pmatrix}.
\end{equation}
The system must be stable for a steady state to exist, to ascertain this, the Routh-Hurwitz criterion \cite{dejesus1987routh} is employed to characterize the stability  of the system. To achieve this,  the real part of the spectrum of the drift matrix $\mathscr{A}$, Eq.~(\ref{driftmatrix}), must be negative, this means that all the eigenvalues of the drift matrix $\mathscr{A}$ have non-positive real parts.  

\section{Entropy production rate}\label{entropy_prod_rate}
The basic thermodynamics principle asserts the entropy of an open system, (classical or quantum) evolves as:
\begin{equation}
\frac{{\rm d} S}{{\rm d}t}=\Pi - \Phi,
\end{equation}
where $\Pi\!\geqslant\!0$ is the irreversible entropy production rate and $\Phi$ is the entropy flow from the system to the reservoir. In thermal equilibrium, the steady states are characterized by ${\rm d} S/{\rm d}t\!=\!\Pi\!=\!\Phi\!=\!0$. However, when the system is connected to multiple reservoirs or being externally driven, it may instead reach a nonequilibrium steady states where ${\rm d}S/{\rm d}t\!=\!0$ but $\Pi\!=\!\Phi\!\geqslant0$. In this nonequilibrium steady state case, the system is characterized by the continuous production of entropy, all of which flows to the reservoir.

We now move to study the entropy production rate from a multipartite system. In analogy with the bipartite case~\cite{brunelli2016irreversibility}, we combine quantum phase-space methods and the Fokker-Planck equation to characterise the irreversible entropy production of quantum systems interacting with reservoirs. In general, the entropy production rate of the quantum system described in Section \ref{model} is given by (see, Appendix \ref{appendix}): 
\begin{equation}
\Pi_{\rm s} :=  \sum_{i=1}^3 2\, \gamma_i \left(\frac{V_{2i-1,2i-1} + V_{2i,2i}}{2\, \mathcal{N}_i + 1} - 1 \right),
\label{gen_entropy}
\end{equation}
where $\Gamma_i \in \{\Gamma_{\rm a}, \Gamma_{\rm m}, \Gamma_{\rm b}\}$ with $\Gamma \in \{\gamma, \mathcal{N}\} $.

Here, we focus on characterizing the entropy production rates in a cavity magnomechanical system. Without loss of generality, we consider the effective entropy flow between the magnon mode and the mechanical resonator. Thus, the rate of entropy production $\Pi_{\rm s}$ at a steady-state reads 
\begin{equation}
\Pi_{\rm s} = 2\gamma_m \left( \mathscr{V}_{33}+\mathscr{V}_{44}-1 \right) +  2\gamma_b \left( \frac{\mathscr{V}_{55} +\mathscr{V}_{66}}{2\mathcal{N}_{\rm b} +1}-1 \right).
\end{equation}
We remark, when the system is in the equilibrium state, we have $\mathscr{V}_{11}+\mathscr{V}_{22} = 1$, $\mathscr{V} _{33}+\mathscr{V}_{44}\!=\!2\mathcal{N}_{\rm b} + 1$, and hence, $\Pi_{\rm s}\!\equiv \! 0$. 

To proceed, we study the entropy production rate in a magnon-phonon-photon system at a steady state in the resolved sideband, where the magnon dissipation rate is comparable to or well below the mechanical resonance frequency  (i.e., $\gamma_{\rm m} < \Omega_{\rm b}$). 
We assume the following parameters close to those employed in the experimental realizations~\cite{zhang2016cavity}, as the phonon frequency $\Omega_{\rm b}/2\pi\!=\!10$ MHz, the cavity dissipation rate $\gamma_{\rm a}/2\pi\!=\!3$ MHz, the magnon dissipation rate $\gamma_{\rm m}/2\pi\!=\!1$ MHz, the phonon damping rate $\gamma_{\rm b}\!=\!300$ Hz, the phonon-magnon coupling $g_{\rm mb}/2\pi\!\simeq\!1$ Hz, and the temperature $T=10 - 100$ mK (i.e., $\mathcal{N}_{\rm b}\simeq200$). In the following analysis, we will utilize dimensionless quantities; that is, the quantities  will be expressed in units of the phonon frequency, $\Omega_{\rm b}$.

\begin{figure}[th]
\includegraphics[width=1.0 \columnwidth]{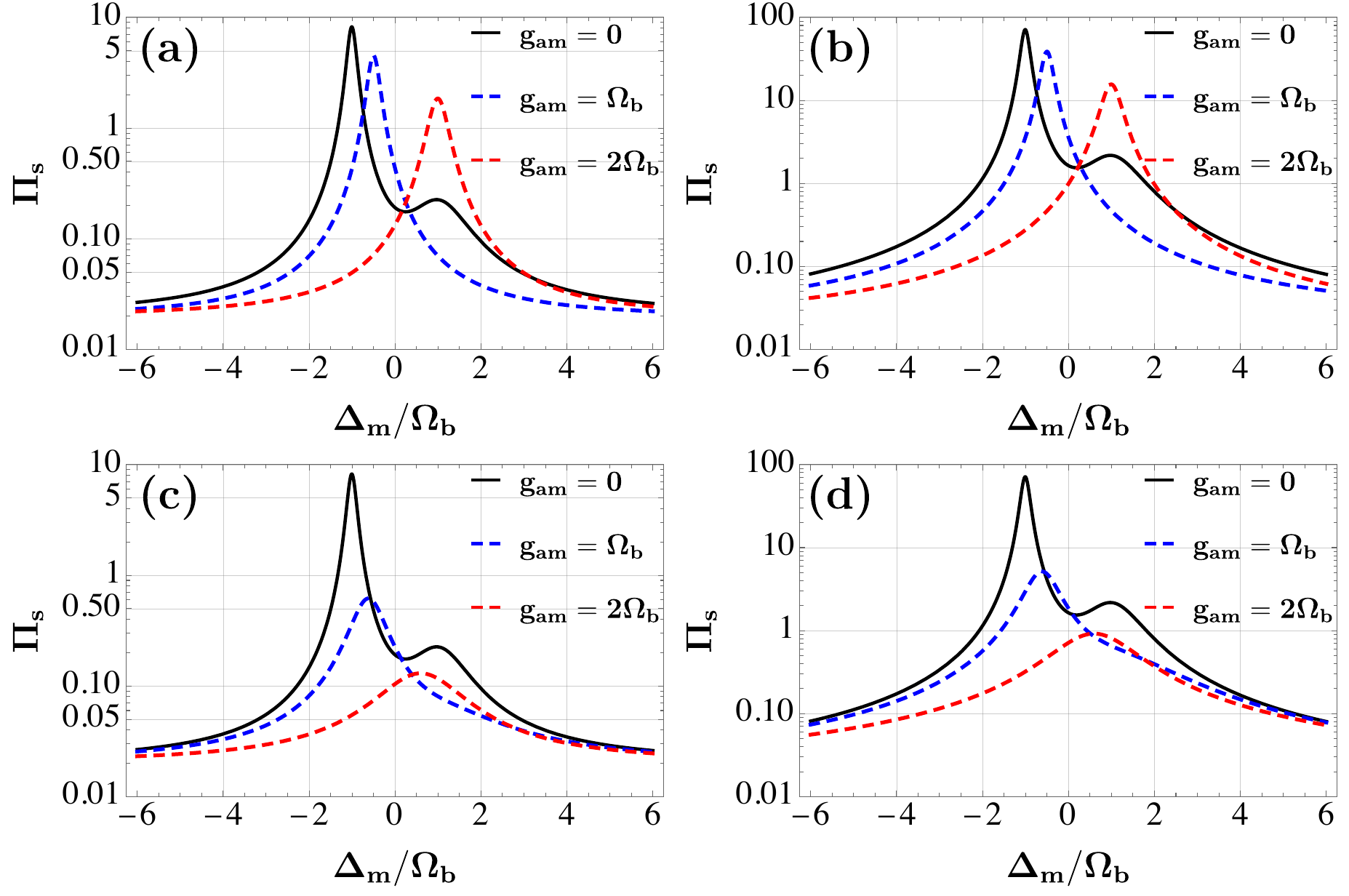} 
\caption{Plot of entropy $\Pi_{\rm s}$ (blue dashed) as a function of normalized magnon detuning $\Delta_{\rm m}/\Omega_{\rm b}$ for different values of magnon-photon coupling $g_{\rm am}\!=\!0$ (black curve), $g_{\rm am}\!=\!\Omega_{\rm b}$ (blue curve) and $g_{\rm am}\!=\!2 \Omega_{\rm b}$ (red curve), where \textbf{(a)} $\mathcal{N}_{\rm b}\!=\!10$ \textbf{(b)} $\mathcal{N}_{\rm b}\!=\!100$,  \textbf{(c)}, $\mathcal{N}_{\rm b}\!=\!10$, and \textbf{(d)}  $\mathcal{N}_{\rm b}\!=\!100$. In \textbf{(a)} \& \textbf{(b)}: $\gamma_{\rm a}\!=\!0.1 \Omega_{\rm b}$, and  $\gamma_{\rm a}\!=\! \Omega_{\rm b}$ in \textbf{(c)} \& \textbf{(d)}. The other parameters are $\mathscr{G}_{\rm mb}\!=\!10^{-1} \Omega_{\rm b}$, $\gamma_{\rm b}=10^{-2}\Omega_{\rm b}$ and $\gamma_{\rm m}\!=\! \Omega_{\rm b}/2$.}         
\label{fig2}
\end{figure}
In Fig. \ref{fig2}, we present the entropy production rate $\Pi_{\rm s}$ is plotted as a function of the normalized magnon detuning $\Delta_{\rm m}/\Omega_{\rm b}$ for various values of the photon-magnon coupling $g_{\rm am}$. In Fig. \ref{fig2}(a) and (b), we consider the dissipation rate of the cavity $\gamma_{\rm a}\!=\! 10^{-1}  \Omega_b$ for distinct occupation number $\mathcal{N}_{\rm b}\!=\!10$ and $\mathcal{N}_{\rm b}\!=\!100$, respectively. It can be seen that  in the absence of magnon-photon interaction $g_{\rm am}\!=\!0$, the entropy production rate $\Pi_{\rm s}$ peaks at $\Delta_{\rm m}/\Omega_{\rm b}\!=\!\pm 1$. The two peaks in the rate of entropy production for positive/negative detuning imply  the cooling/heating processes behave differently in two distinct regimes of the system. For non-zero $g_{\rm am}$, the smaller peak smears out with reduced maximum $\Pi_{\rm s}$. We observe that for $g_{\rm am}\!=\!\Omega_{\rm b}$ ($g_{\rm am}\!=\!2\, \Omega_{\rm b}$), the maximum $\Pi_{\rm s}$ is in the red (blue) sideband region. The maximum entropy flow between the phonon mode and the effective magnon mode occurs when $g_{\rm am}\!=\!2\Omega_b$, at $\Delta_{\rm m}=\omega_{\rm b}$. Fig~\ref{fig2} (c) and (d) show more clearly the effect and interplay between the cavity photon dissipation rate and the magnon-photon coupling. For the case of $\gamma_{\rm b}\!=\!\Omega_{\rm b}$, Fig. \ref{fig2}(c) and (d), the value of $\Pi_{\rm s}$ further decreases with a broaden peak as magnon-phonon coupling $g_{\rm am}$ increases. In the blue-sideband, for $g_{\rm am}\ne 0$, the value of $\Pi_{\rm s}$ is always reduced compare to the $g_{\rm am}\!=\!0$ scenario. It can be seen that $\Pi_{\rm s}$ could be enhanced close to the negative detuning region $\Delta_{\rm m}<0$. The effects of thermal fluctuations of the environment on the entropy production rate is shown in Fig.~\ref{fig2}.
 The Fig. \ref{fig2} shows that increasing the phonon thermal excitation (occupation number) $\mathcal{N}_{\rm b}$, increases $\Pi_{\rm s}$ in magnitude (see the magnitude of $\Pi_{\rm s}$ in Fig.~\ref{fig2}(a) and (c) compare to (b) and (d) respectively). The non-uniform behaviour when varying $g_{\rm am}$ for different $\gamma_{\rm a}$ is a direct implication of the complex nature of the interaction between the dissipation processes in the cavity magnomechanical system.

\begin{figure}[tb]
\includegraphics[width=0.96 \columnwidth]{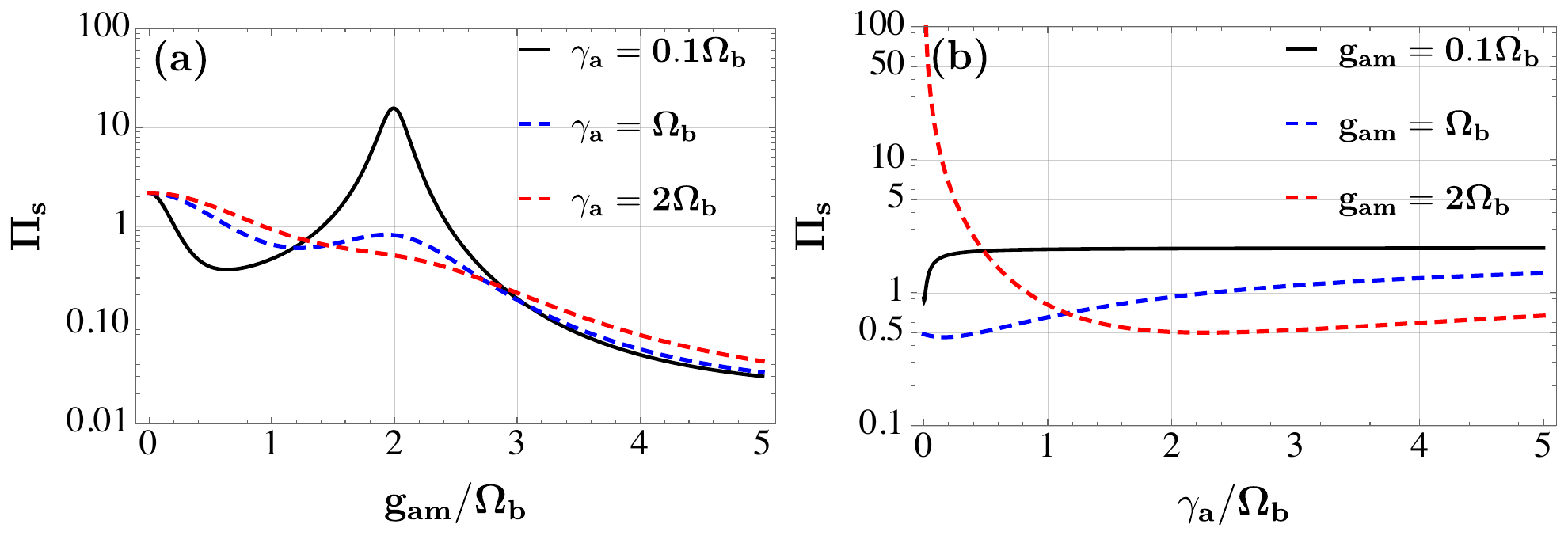} 
\caption{ \textbf{(a)} The entropy production rate $\Pi_{\rm s}$ as a function of magnon-photon coupling $g_{\rm am}/\Omega_{\rm b}$ for different values of cavity photon decay rate $\gamma_{\rm a}\!=\!  10^{-1}  \Omega_{\rm b}$ (black curve), $\gamma_{\rm a}\!=\! \Omega_{\rm b}$ (blue curve) and $\gamma_{\rm a}\!=\!2\, \Omega_{\rm b}$ (red curve). \textbf{(b)} Entropy production rate  $\Pi_{\rm s}$ as a function of cavity photon decay rate $\gamma_{\rm a}/\Omega_{\rm b}$ for different values of magnon-photon coupling $g_{\rm am}\!=\!  10^{-1}  \, \Omega_{\rm b}$ (black curve), $g_{\rm am}\!=\! \Omega_{\rm b}$ (blue curve) and $g_{\rm am}=2\, \Omega_{\rm b}$ (red curve). Other parameters are  $\mathcal{N}_{\rm b}=100$, $\Delta_{\rm m}=\Omega_{\rm b}$, $\mathscr{G}_{\rm mb}\! =\! 10^{-1} \Omega_{\rm b}$, $\gamma_{\rm b}\!=\!10^{-2}\Omega_{\rm b}$ and $\gamma_{\rm m}\!=\!\Omega_{\rm b}/2$.}         
\label{fig3}
\end{figure}

In Fig.~\ref{fig3}, the effects of the cavity photon dissipation rate of the environment on the entropy production rate is explored.
We present in Fig.\ref{fig3} (a),  the plot of the entropy production rate as a function of magnon-photon coupling with various values of the cavity decay rate, while Fig.~\ref{fig3} (b)  shows  the entropy production rate as function of cavity decay rate  with different values of the  magnon-photon coupling. For large thermal excitations, $\mathcal{N}_{\rm b}\!=\!100$, and $\Delta_{\rm m}\!=\!\Omega_{\rm b}$, the entropy production rates decrease in oscillatory form as the magnon-photon coupling increases. For $\gamma_{\rm a}\!=\!  10^{-1}  \, \Omega_{\rm b}$, the maximum entropy production rate corresponds to the $g_{\rm am}\!=\!2 \Omega_{\rm b}$. It can be seen that the entropy production rate can be increase/decrease by tuning the cavity decay value  $\gamma_{\rm a}$ and the magnon-photon coupling, see $g_{\rm am}\!\simeq\!0.2 - 3\, \Omega_b$. The entropy production rate  $\Pi_{\rm s}$ linearly decreases with respect to the magnon-photon coupling when $g_{\rm am}>3/\Omega_{\rm b}$ but the value slightly increase when we increase $\gamma_{\rm a}$. This non-monotonic behaviour (increase/decrease) of the entropy production with respect to the magnon-photon coupling can be attributed to the imbalance in populations between the interacting modes induced by the environment. Furthermore, Fig.~\ref{fig3} (b) shows that the entropy production rate can increase/decrease when $\gamma_{\rm a}\!<\!g_{\rm am}$ but increases slightly after $\gamma_{\rm a}\!=\!g_{\rm am}$. In the region, $\gamma_{\rm a} \gg g_{\rm am}$, the entropy production rate saturates to a quasi-constant value with decreasing value as the magnon-photon $g_{\rm am}$ increases.
 This is due to the fact that the individual modes are far from resonance and they are effectively decoupled, in such a way that the individually thermalize their own bath. 

\begin{figure*}[!]
\includegraphics[width=2.1\columnwidth]{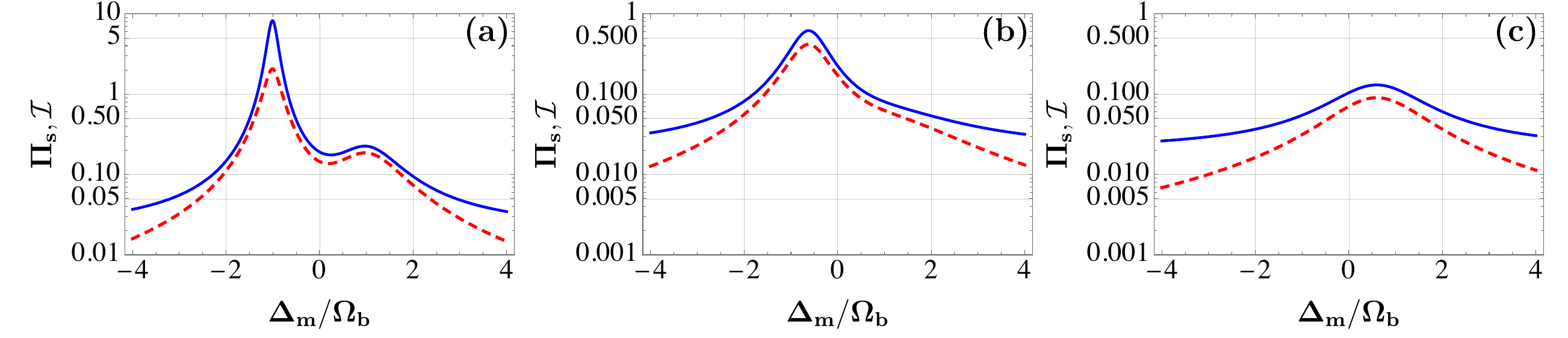} 
\caption{Entropy rate $\Pi_{\rm s}$ (blue curve) and mutual information $\mathcal{I}$ (red curve) as a function of normalized magnon detuning $\Delta_{\rm m}/\Omega_{\rm b}$ for different values of magnon-photon coupling (a) $g_{\rm am}\!=\!0$, (b) $g_{\rm am}\!=\!\Omega_{\rm b}$ and (c) $g_{\rm am}\!=\!2\, \Omega_{\rm b}$. Here, other parameters are $\gamma_{\rm a}=\Omega_{\rm b}$, $\mathcal{N}_{\rm b}=10$, $\mathscr{G}_{\rm mb}\!= \! 10^{-1}  \Omega_{\rm b}$, $\gamma_{\rm b}=10^{-2}\Omega_{\rm b}$ and $\gamma_{\rm m}= \Omega_{\rm b}/2$.}         
\label{fig4}
\end{figure*}
Next, we analyze the behaviour of the entropy production rate with respect to the amount of correlations shared between the magnon and phonon modes. For coupled quantum system, it has   been demonstrated that the irreversibility generated by the dissipative system at steady state and the total amount of correlations shared between the subsystems are closely related  in small coupling limit as \cite{brunelli2016irreversibility}; $\mathcal{I}\! \approx \! \Pi_{\rm s} /(2\,\gamma_\text{tot})$, where $\mathcal{I}$ is the quantum mutual information between the modes at the stationary state and $\gamma_\text{tot}$ is the sum of the dissipation rates to the local baths. 
Since the quantum noises are Gaussian, the mutual information between the magnon and phonon modes can be computed as \cite{adesso2012measuring}:
\begin{equation}
    \mathcal{I}(\mathcal{V}_{\rm a:b}) = \frac{1}{2} \ln \left(\frac{\text{det}(\mathcal{V}_{\rm a}) \, \text{det}(\mathcal{V}_{\rm b})}{\text{det} (\mathcal{V}_{\rm ab})}\right),
\end{equation}
where $\mathcal{V}_{\rm a}$ $\left(\mathcal{V}_{\rm b}\right)$ is the covariance matrix of magnon (mechanical) mode.

In Fig. \ref{fig4}, we show a comparative plot of the entropy production rate $\Pi_S$ to the correlations established by the cavity magnomechanical system, as quantified by the mutual information $\mathcal{I}$. 
Considering small thermal phonon excitation, $\mathcal{N}_{\rm b}\!=\!10$, Fig.~\ref{fig4} (a) shows a good similarity in both $\Pi_{\rm s}$ and $\mathcal{I}$ curves for $|\Delta_{\rm m}/\Omega_{\rm b}|\leqslant 2$. As the magnon-photon coupling increases, both the entropy production rate and mutual information are decreasing as well as an increase in both quantities deviation, see Fig. \ref{fig4} (b) and (c). The increase in deviation between $\Pi_{\rm s}$ and $\mathcal{I}$ for increasing detuning $\Delta_{\rm m}$, clearly demonstrates the interplay among magnon-photon coupling and dissipation rates. This shows that degree of irreversibility induced by the stationary process is directly related to the amount of correlations shared in the system.

\section{Conclusions}\label{conclusion}
We have investigated the entropy production rate in a hybrid magnomechanical system where a microwave cavity mode is coupled to a magnon mode in a YIG sphere, and the latter is simultaneously coupled to a phonon mode via magnetostrictive force. Specifically, within the quantum phase space formulation of the entropy change, we evaluate the steady-state entropy production rate and associated quantum correlation in the hybrid system.  We have shown that the entropy flow between the effective magnon mode and the phonon mode is influenced by the magnon-photon coupling and the cavity photon dissipation rate. Our numerical analysis shows non-uniform behavior of irreversibility resulting from the complex and competing nature of the interactions in the cavity magnomechanical system. We find that the entropy production rate can be increased/decreased when the cavity decay rate is less than the magnon-photon coupling. Furthermore, we studied the range of validity of the link between irreversibility and the amount of correlations in a mesoscopic quantum system. These results provide insight into the impact of magnonics on the thermodynamics processes of hybrid quantum systems. Finally, we anticipate that our study will open new perspectives for quantum thermodynamics applications, such as realizing thermal machines in the deep quantum regime and quantum thermal transport.

\section*{Acknowledgement}
COE and NA have been supported by LRGS Grant LRGS/1/2020/UM/01/5/2 (9012-00009) provided by the Ministry of Higher Education of Malaysia (MOHE). MA and DD have been supported by the Khalifa University of Science and Technology under Award
No. FSU- 2023-014. 

\appendix
\section{Entropy production}\label{appendix}
In this appendix, we provide the explicit expression for the entropy production rate presented in Eq.~\ref{gen_entropy}. Considering the Gaussian nature of the quantum system, their states are characterized by the Wigner function of the form
\begin{equation}
\mathcal{W}(R)=\frac{1}{\pi^n \sqrt{\text{det}\, \sigma}} \exp \left(- \frac{1}{2} (R-\bar{R})^\top \sigma^{-1} (R-\bar{R}) \right),
\end{equation}
where $n$ is the number of the bosonic mode and $\sigma$ is the  covariance matrix (CM), whose entries are given by $\sigma_{ij}\!=\! \frac{1}{2}\langle {R_i R_j+R_j R_i}\rangle - \langle R_i\rangle \langle R_j\rangle$. The entropy of the Gaussian system is associated to the corresponding Shannon entropy (Wigner entropy) of the Wigner distribution.  

Following the quantum Langenvin equations for the system dynamics described in Section \ref{model}, we can recast the Fokker-Planck equation for the Wigner function $\mathcal{W}(u,t)$  as a local conservation equation
\begin{equation}
\partial_t\mathcal{W} = -\partial_u \mathcal{J}(u,t),
\end{equation}
where $u=(x_a,y_a,x_m,y_m,x_b,x_b)^\top$ is a point in phase space, $\partial_u$ is the phase-space gradient, and $\mathcal{J}$ is the total probability current vector that reads
\begin{equation}
\mathcal{J}(u,t)\!=\!\mathscr{A} u {W}(u,t) - \frac{1}{2} \mathscr{D} \partial_u\mathcal{W}(u,t).
\end{equation} 
For the three mode bosonic system model that we consider, the drift matrix $\mathscr{A}$ and the diffusion matrix $\mathscr{D}$ are defined  in Section \ref{model} (Eq. \ref{driftmatrix}). Then, introducing the time-reversal operator $E= \text{diag} (1, -1, 1, -1, 1, -1)$, the dynamical variables can be split according to their symmetry. The drift matrix $\mathscr{A}$ can be rewritten by splitting it into the irreversible part $\mathscr{A}^{\text{irr}}$ which is even under time reversal and the reversible part $\mathscr{A}^{\text{rev}}$ that is odd \cite{brunelli2016irreversibility}. They can be evaluated as $\mathscr{A}^{\text{irr}}:=\frac{1}{2}(\mathscr{A} + E \mathscr{A} E^\top)$ and $\mathscr{A}^{\text{rev}}\!:=\!\frac{1}{2}(\mathscr{A} - E \mathscr{A} E^\top)$. The irreversible part is associated with the damping rates, while the reversible part comes from the Hamiltonian part of the dynamics. The drift matrix separation results in splitting of the probability currents as $\mathcal{J}(u,t) \equiv \mathcal{J}^{\text{rev}}(u,t)+\mathcal{J}^{\text{irr}}(u,t)$, where
\begin{equation}
\mathcal{J}^{\text{rev}}(u,t) := \mathscr{A}^{\text{rev}} u W(u,t),
\end{equation}
and 
\begin{equation}
\mathcal{J}^{\text{irr}}(u,t) := \mathscr{A}^{\text{irr}} u W(u,t) - \frac{1}{2} \mathscr{D} \partial_u W(u,t).
\end{equation}

Taking the derivative of the Wigner entropy with respect to time and integrating by parts, we get
\begin{equation}
S_W = \int d^2 \alpha (\mathcal{J}^{\text{irr}} \mathcal{W})^\top \left(\frac{\partial_u \mathcal{W}}{\mathcal{W}}\right).
\label{Wentropy}
\end{equation}
Equation (\ref{Wentropy}) can be rewritten in the usual form in terms of the entropy production rate \cite{brunelli2016irreversibility}:
\begin{equation}
\Pi= \frac{1}{2} \int \text{d}u \, \frac{1}{\mathcal{W}(u,t)} \mathcal{J}^{\text{irr}}(u,t)^\top  D^{-1} \mathcal{J}^{\text{irr}}(u,t),
\label{gen_Pi}
\end{equation}
 and entropy flux rate
\begin{equation}
\Phi = -\frac{1}{2} \int \text{d}u \, \mathcal{J}^{\text{irr}}(u,t) D^{-1} A^{\text{irr}} u.
\end{equation}
Moreover, for a Gaussian state, $\partial_u \mathcal{W}_\sigma (u,t) = -\mathcal{W}_\sigma (u,t) \sigma^{-1}(t) u$, then the irreversible component of probability current $\mathcal{J}^{\text{irr}}(u,t)=\mathcal{W}_\sigma (u,t) (\mathscr{A}^{\text{irr}}u + D) \sigma^{-1}(t) u$. Therefore, the explicit integration of Eq. (\ref{gen_Pi}) gives the entropy production rate as \cite{brunelli2016irreversibility}
\begin{equation}
\Pi = \frac{1}{2} {\rm tr} \left[\sigma^{-1} \mathscr{D}\right] + 2 \mathrm{tr} [\mathscr{A}^\text{irr}] + 2 \mathrm{tr} \left[ (\mathscr{A}^\text{irr})^\top  \mathscr{D}^{-1}  \mathscr{A}^\text{irr}\, \sigma  \right].
\end{equation}
For the case of three mode bosonic system, in stationary state, we obtain
\begin{eqnarray}
\Pi_{\rm s} &=&  2\gamma_a \left( \frac{\mathscr{V}_{11} +\mathscr{V}_{22}}{2\mathcal{N}_{\rm a} +1}-1 \right) +  2\gamma_m \left( \frac{\mathscr{V}_{33} +\mathscr{V}_{44}}{2\mathcal{N}_{\rm m} +1}-1 \right)  \nonumber\\ 
&+&  2\gamma_b \left( \frac{\mathscr{V}_{55} +\mathscr{V}_{66}}{2\mathcal{N}_{\rm b} +1}-1 \right) \\
&=& \sum_{i=1}^3 2\, \gamma_i \left(\frac{V_{2i-1,2i-1} + V_{2i,2i}}{2\, \mathcal{N}_i + 1} - 1 \right)
\end{eqnarray}
where  $\Gamma_i \in \{\Gamma_{\rm a}, \Gamma_{\rm m}, \Gamma_{\rm b}\}$ with $\Gamma \in \{\gamma, \mathcal{N}\} $.

Within this framework, the rate of entropy production $\Pi_{\rm s}$ at steady-state for two coupled oscillators mode is expressed as follows \cite{brunelli2016irreversibility,brunelli2018experimental},
\begin{eqnarray}
\Pi_{\rm s} &=& \mathrm{Tr} \left(2  \mathscr{A}^\text{irr}  \mathscr{D}^{-1}  \mathscr{A}^\text{irr}\, \mathscr{V} +  \mathscr{A}^\text{irr} \right) \nonumber\\
&=& 2\gamma_m \left( \mathscr{V}_{33}+\mathscr{V}_{44}-1 \right) +  2\gamma_b \left( \frac{\mathscr{V}_{55} +\mathscr{V}_{66}}{2\mathcal{N}_{\rm b} +1}-1 \right) ,\nonumber \\
\end{eqnarray}
where $ \mathscr{A}^\text{irr} = \text{diag}\{0, 0, -\gamma_{\rm m},-\gamma_{\rm m},-\gamma_{\rm b},-\gamma_{\rm b}\}$. 
This is the expression given in the main text.

\begin{thebibliography}{53}%
\makeatletter
\providecommand \@ifxundefined [1]{%
 \@ifx{#1\undefined}
}%
\providecommand \@ifnum [1]{%
 \ifnum #1\expandafter \@firstoftwo
 \else \expandafter \@secondoftwo
 \fi
}%
\providecommand \@ifx [1]{%
 \ifx #1\expandafter \@firstoftwo
 \else \expandafter \@secondoftwo
 \fi
}%
\providecommand \natexlab [1]{#1}%
\providecommand \enquote  [1]{``#1''}%
\providecommand \bibnamefont  [1]{#1}%
\providecommand \bibfnamefont [1]{#1}%
\providecommand \citenamefont [1]{#1}%
\providecommand \href@noop [0]{\@secondoftwo}%
\providecommand \href [0]{\begingroup \@sanitize@url \@href}%
\providecommand \@href[1]{\@@startlink{#1}\@@href}%
\providecommand \@@href[1]{\endgroup#1\@@endlink}%
\providecommand \@sanitize@url [0]{\catcode `\\12\catcode `\$12\catcode
  `\&12\catcode `\#12\catcode `\^12\catcode `\_12\catcode `\%12\relax}%
\providecommand \@@startlink[1]{}%
\providecommand \@@endlink[0]{}%
\providecommand \url  [0]{\begingroup\@sanitize@url \@url }%
\providecommand \@url [1]{\endgroup\@href {#1}{\urlprefix }}%
\providecommand \urlprefix  [0]{URL }%
\providecommand \Eprint [0]{\href }%
\providecommand \doibase [0]{https://doi.org/}%
\providecommand \selectlanguage [0]{\@gobble}%
\providecommand \bibinfo  [0]{\@secondoftwo}%
\providecommand \bibfield  [0]{\@secondoftwo}%
\providecommand \translation [1]{[#1]}%
\providecommand \BibitemOpen [0]{}%
\providecommand \bibitemStop [0]{}%
\providecommand \bibitemNoStop [0]{.\EOS\space}%
\providecommand \EOS [0]{\spacefactor3000\relax}%
\providecommand \BibitemShut  [1]{\csname bibitem#1\endcsname}%
\let\auto@bib@innerbib\@empty
\bibitem [{\citenamefont {Xiang}\ \emph {et~al.}(2013)\citenamefont {Xiang},
  \citenamefont {Ashhab}, \citenamefont {You},\ and\ \citenamefont
  {Nori}}]{xiang2013hybrid}%
  \BibitemOpen
  \bibfield  {author} {\bibinfo {author} {\bibfnamefont {Z.-L.}\ \bibnamefont
  {Xiang}}, \bibinfo {author} {\bibfnamefont {S.}~\bibnamefont {Ashhab}},
  \bibinfo {author} {\bibfnamefont {J.~Q.}\ \bibnamefont {You}},\ and\ \bibinfo
  {author} {\bibfnamefont {F.}~\bibnamefont {Nori}},\ }\bibfield  {title}
  {\bibinfo {title} {Hybrid quantum circuits: Superconducting circuits
  interacting with other quantum systems},\ }\href
  {https://doi.org/10.1103/RevModPhys.85.623} {\bibfield  {journal} {\bibinfo
  {journal} {Rev. Mod. Phys.}\ }\textbf {\bibinfo {volume} {85}},\ \bibinfo
  {pages} {623} (\bibinfo {year} {2013})}\BibitemShut {NoStop}%
\bibitem [{\citenamefont {Bauer}\ \emph {et~al.}(2016)\citenamefont {Bauer},
  \citenamefont {Wecker}, \citenamefont {Millis}, \citenamefont {Hastings},\
  and\ \citenamefont {Troyer}}]{bauer2016hybrid}%
  \BibitemOpen
  \bibfield  {author} {\bibinfo {author} {\bibfnamefont {B.}~\bibnamefont
  {Bauer}}, \bibinfo {author} {\bibfnamefont {D.}~\bibnamefont {Wecker}},
  \bibinfo {author} {\bibfnamefont {A.~J.}\ \bibnamefont {Millis}}, \bibinfo
  {author} {\bibfnamefont {M.~B.}\ \bibnamefont {Hastings}},\ and\ \bibinfo
  {author} {\bibfnamefont {M.}~\bibnamefont {Troyer}},\ }\bibfield  {title}
  {\bibinfo {title} {Hybrid quantum-classical approach to correlated
  materials},\ }\href {https://doi.org/10.1103/PhysRevX.6.031045} {\bibfield
  {journal} {\bibinfo  {journal} {Phys. Rev. X}\ }\textbf {\bibinfo {volume}
  {6}},\ \bibinfo {pages} {031045} (\bibinfo {year} {2016})}\BibitemShut
  {NoStop}%
\bibitem [{\citenamefont {Li}\ \emph {et~al.}(2017)\citenamefont {Li},
  \citenamefont {Yang}, \citenamefont {Peng},\ and\ \citenamefont
  {Sun}}]{lihybrid:2017}%
  \BibitemOpen
  \bibfield  {author} {\bibinfo {author} {\bibfnamefont {J.}~\bibnamefont
  {Li}}, \bibinfo {author} {\bibfnamefont {X.}~\bibnamefont {Yang}}, \bibinfo
  {author} {\bibfnamefont {X.}~\bibnamefont {Peng}},\ and\ \bibinfo {author}
  {\bibfnamefont {C.-P.}\ \bibnamefont {Sun}},\ }\bibfield  {title} {\bibinfo
  {title} {Hybrid quantum-classical approach to quantum optimal control},\
  }\href {https://doi.org/10.1103/PhysRevLett.118.150503} {\bibfield  {journal}
  {\bibinfo  {journal} {Phys. Rev. Lett.}\ }\textbf {\bibinfo {volume} {118}},\
  \bibinfo {pages} {150503} (\bibinfo {year} {2017})}\BibitemShut {NoStop}%
\bibitem [{\citenamefont {Lachance-Quirion}\ \emph {et~al.}(2019)\citenamefont
  {Lachance-Quirion}, \citenamefont {Tabuchi}, \citenamefont {Gloppe},
  \citenamefont {Usami},\ and\ \citenamefont {Nakamura}}]{lachance2019hybrid}%
  \BibitemOpen
  \bibfield  {author} {\bibinfo {author} {\bibfnamefont {D.}~\bibnamefont
  {Lachance-Quirion}}, \bibinfo {author} {\bibfnamefont {Y.}~\bibnamefont
  {Tabuchi}}, \bibinfo {author} {\bibfnamefont {A.}~\bibnamefont {Gloppe}},
  \bibinfo {author} {\bibfnamefont {K.}~\bibnamefont {Usami}},\ and\ \bibinfo
  {author} {\bibfnamefont {Y.}~\bibnamefont {Nakamura}},\ }\bibfield  {title}
  {\bibinfo {title} {Hybrid quantum systems based on magnonics},\ }\href
  {https://doi.org/https://doi.org/10.7567/1882-0786/ab248d} {\bibfield
  {journal} {\bibinfo  {journal} {Appl. Phys. Express}\ }\textbf {\bibinfo
  {volume} {12}},\ \bibinfo {pages} {070101} (\bibinfo {year}
  {2019})}\BibitemShut {NoStop}%
\bibitem [{\citenamefont {Clerk}\ \emph {et~al.}(2020)\citenamefont {Clerk},
  \citenamefont {Lehnert}, \citenamefont {Bertet}, \citenamefont {Petta},\ and\
  \citenamefont {Nakamura}}]{clerk2020hybrid}%
  \BibitemOpen
  \bibfield  {author} {\bibinfo {author} {\bibfnamefont {A.}~\bibnamefont
  {Clerk}}, \bibinfo {author} {\bibfnamefont {K.}~\bibnamefont {Lehnert}},
  \bibinfo {author} {\bibfnamefont {P.}~\bibnamefont {Bertet}}, \bibinfo
  {author} {\bibfnamefont {J.}~\bibnamefont {Petta}},\ and\ \bibinfo {author}
  {\bibfnamefont {Y.}~\bibnamefont {Nakamura}},\ }\bibfield  {title} {\bibinfo
  {title} {Hybrid quantum systems with circuit quantum electrodynamics},\
  }\href {https://doi.org/https://doi.org/10.1038/s41567-020-0797-9} {\bibfield
   {journal} {\bibinfo  {journal} {Nat. Phys.}\ }\textbf {\bibinfo {volume}
  {16}},\ \bibinfo {pages} {257} (\bibinfo {year} {2020})}\BibitemShut
  {NoStop}%
\bibitem [{\citenamefont {Callison}\ and\ \citenamefont
  {Chancellor}(2022)}]{callison2022hybrid}%
  \BibitemOpen
  \bibfield  {author} {\bibinfo {author} {\bibfnamefont {A.}~\bibnamefont
  {Callison}}\ and\ \bibinfo {author} {\bibfnamefont {N.}~\bibnamefont
  {Chancellor}},\ }\bibfield  {title} {\bibinfo {title} {Hybrid
  quantum-classical algorithms in the noisy intermediate-scale quantum era and
  beyond},\ }\href
  {https://doi.org/https://doi.org/10.1103/PhysRevA.106.010101} {\bibfield
  {journal} {\bibinfo  {journal} {Phys. Rev. A}\ }\textbf {\bibinfo {volume}
  {106}},\ \bibinfo {pages} {010101} (\bibinfo {year} {2022})}\BibitemShut
  {NoStop}%
\bibitem [{\citenamefont {Potter}\ and\ \citenamefont
  {Vasseur}(2022)}]{potter2022entanglement}%
  \BibitemOpen
  \bibfield  {author} {\bibinfo {author} {\bibfnamefont {A.~C.}\ \bibnamefont
  {Potter}}\ and\ \bibinfo {author} {\bibfnamefont {R.}~\bibnamefont
  {Vasseur}},\ }\bibfield  {title} {\bibinfo {title} {Entanglement dynamics in
  hybrid quantum circuits},\ }in\ \href
  {https://doi.org/https://doi.org/10.1007/978-3-031-03998-0_9} {\emph
  {\bibinfo {booktitle} {Entanglement in Spin Chains}}}\ (\bibinfo  {publisher}
  {Springer},\ \bibinfo {year} {2022})\ pp.\ \bibinfo {pages}
  {211--249}\BibitemShut {NoStop}%
\bibitem [{\citenamefont {Wallquist}\ \emph {et~al.}(2009)\citenamefont
  {Wallquist}, \citenamefont {Hammerer}, \citenamefont {Rabl}, \citenamefont
  {Lukin},\ and\ \citenamefont {Zoller}}]{wallquist2009hybrid}%
  \BibitemOpen
  \bibfield  {author} {\bibinfo {author} {\bibfnamefont {M.}~\bibnamefont
  {Wallquist}}, \bibinfo {author} {\bibfnamefont {K.}~\bibnamefont {Hammerer}},
  \bibinfo {author} {\bibfnamefont {P.}~\bibnamefont {Rabl}}, \bibinfo {author}
  {\bibfnamefont {M.}~\bibnamefont {Lukin}},\ and\ \bibinfo {author}
  {\bibfnamefont {P.}~\bibnamefont {Zoller}},\ }\bibfield  {title} {\bibinfo
  {title} {Hybrid quantum devices and quantum engineering},\ }\href
  {https://doi.org/https://doi.org/10.1088/0031-8949/2009/T137/014001}
  {\bibfield  {journal} {\bibinfo  {journal} {Phys. Scr.}\ }\textbf {\bibinfo
  {volume} {2009}},\ \bibinfo {pages} {014001} (\bibinfo {year}
  {2009})}\BibitemShut {NoStop}%
\bibitem [{\citenamefont {Degen}\ \emph {et~al.}(2017)\citenamefont {Degen},
  \citenamefont {Reinhard},\ and\ \citenamefont {Cappellaro}}]{Degen2017}%
  \BibitemOpen
  \bibfield  {author} {\bibinfo {author} {\bibfnamefont {C.~L.}\ \bibnamefont
  {Degen}}, \bibinfo {author} {\bibfnamefont {F.}~\bibnamefont {Reinhard}},\
  and\ \bibinfo {author} {\bibfnamefont {P.}~\bibnamefont {Cappellaro}},\
  }\bibfield  {title} {\bibinfo {title} {Quantum sensing},\ }\href
  {https://doi.org/10.1103/RevModPhys.89.035002} {\bibfield  {journal}
  {\bibinfo  {journal} {Rev. Mod. Phys.}\ }\textbf {\bibinfo {volume} {89}},\
  \bibinfo {pages} {035002} (\bibinfo {year} {2017})}\BibitemShut {NoStop}%
\bibitem [{\citenamefont {Myers}\ \emph {et~al.}(2022)\citenamefont {Myers},
  \citenamefont {Abah},\ and\ \citenamefont {Deffner}}]{myers2022quantum}%
  \BibitemOpen
  \bibfield  {author} {\bibinfo {author} {\bibfnamefont {N.~M.}\ \bibnamefont
  {Myers}}, \bibinfo {author} {\bibfnamefont {O.}~\bibnamefont {Abah}},\ and\
  \bibinfo {author} {\bibfnamefont {S.}~\bibnamefont {Deffner}},\ }\bibfield
  {title} {\bibinfo {title} {Quantum thermodynamic devices: from theoretical
  proposals to experimental reality},\ }\href
  {https://doi.org/https://doi.org/10.1116/5.0083192} {\bibfield  {journal}
  {\bibinfo  {journal} {AVS Quantum Science}\ }\textbf {\bibinfo {volume}
  {4}},\ \bibinfo {pages} {027101} (\bibinfo {year} {2022})}\BibitemShut
  {NoStop}%
\bibitem [{\citenamefont {Aspelmeyer}\ \emph {et~al.}(2014)\citenamefont
  {Aspelmeyer}, \citenamefont {Kippenberg},\ and\ \citenamefont
  {Marquardt}}]{aspelmeyerbook:2014}%
  \BibitemOpen
  \bibfield  {author} {\bibinfo {author} {\bibfnamefont {M.}~\bibnamefont
  {Aspelmeyer}}, \bibinfo {author} {\bibfnamefont {T.~J.}\ \bibnamefont
  {Kippenberg}},\ and\ \bibinfo {author} {\bibfnamefont {F.}~\bibnamefont
  {Marquardt}},\ }\href
  {https://link.springer.com/book/10.1007/978-3-642-55312-7} {\emph {\bibinfo
  {title} {Cavity optomechanics: nano-and micromechanical resonators
  interacting with light}}}\ (\bibinfo  {publisher} {Springer},\ \bibinfo
  {year} {2014})\BibitemShut {NoStop}%
\bibitem [{\citenamefont {Zhang}\ \emph {et~al.}(2016)\citenamefont {Zhang},
  \citenamefont {Zou}, \citenamefont {Jiang},\ and\ \citenamefont
  {Tang}}]{zhang2016cavity}%
  \BibitemOpen
  \bibfield  {author} {\bibinfo {author} {\bibfnamefont {X.}~\bibnamefont
  {Zhang}}, \bibinfo {author} {\bibfnamefont {C.-L.}\ \bibnamefont {Zou}},
  \bibinfo {author} {\bibfnamefont {L.}~\bibnamefont {Jiang}},\ and\ \bibinfo
  {author} {\bibfnamefont {H.~X.}\ \bibnamefont {Tang}},\ }\bibfield  {title}
  {\bibinfo {title} {Cavity magnomechanics},\ }\href
  {https://doi.org/https://doi.org/10.1126/sciadv.1501286} {\bibfield
  {journal} {\bibinfo  {journal} {Sci. Adv.}\ }\textbf {\bibinfo {volume}
  {2}},\ \bibinfo {pages} {e1501286} (\bibinfo {year} {2016})}\BibitemShut
  {NoStop}%
\bibitem [{\citenamefont {Tabuchi}\ \emph {et~al.}(2014)\citenamefont
  {Tabuchi}, \citenamefont {Ishino}, \citenamefont {Ishikawa}, \citenamefont
  {Yamazaki}, \citenamefont {Usami},\ and\ \citenamefont
  {Nakamura}}]{Tabuchi2014}%
  \BibitemOpen
  \bibfield  {author} {\bibinfo {author} {\bibfnamefont {Y.}~\bibnamefont
  {Tabuchi}}, \bibinfo {author} {\bibfnamefont {S.}~\bibnamefont {Ishino}},
  \bibinfo {author} {\bibfnamefont {T.}~\bibnamefont {Ishikawa}}, \bibinfo
  {author} {\bibfnamefont {R.}~\bibnamefont {Yamazaki}}, \bibinfo {author}
  {\bibfnamefont {K.}~\bibnamefont {Usami}},\ and\ \bibinfo {author}
  {\bibfnamefont {Y.}~\bibnamefont {Nakamura}},\ }\bibfield  {title} {\bibinfo
  {title} {Hybridizing ferromagnetic magnons and microwave photons in the
  quantum limit},\ }\href {https://doi.org/10.1103/PhysRevLett.113.083603}
  {\bibfield  {journal} {\bibinfo  {journal} {Phys. Rev. Lett.}\ }\textbf
  {\bibinfo {volume} {113}},\ \bibinfo {pages} {083603} (\bibinfo {year}
  {2014})}\BibitemShut {NoStop}%
\bibitem [{\citenamefont {Bai}\ \emph {et~al.}(2015)\citenamefont {Bai},
  \citenamefont {Harder}, \citenamefont {Chen}, \citenamefont {Fan},
  \citenamefont {Xiao},\ and\ \citenamefont {Hu}}]{Bai2015}%
  \BibitemOpen
  \bibfield  {author} {\bibinfo {author} {\bibfnamefont {L.}~\bibnamefont
  {Bai}}, \bibinfo {author} {\bibfnamefont {M.}~\bibnamefont {Harder}},
  \bibinfo {author} {\bibfnamefont {Y.~P.}\ \bibnamefont {Chen}}, \bibinfo
  {author} {\bibfnamefont {X.}~\bibnamefont {Fan}}, \bibinfo {author}
  {\bibfnamefont {J.~Q.}\ \bibnamefont {Xiao}},\ and\ \bibinfo {author}
  {\bibfnamefont {C.-M.}\ \bibnamefont {Hu}},\ }\bibfield  {title} {\bibinfo
  {title} {Spin pumping in electrodynamically coupled magnon-photon systems},\
  }\href {https://doi.org/10.1103/PhysRevLett.114.227201} {\bibfield  {journal}
  {\bibinfo  {journal} {Phys. Rev. Lett.}\ }\textbf {\bibinfo {volume} {114}},\
  \bibinfo {pages} {227201} (\bibinfo {year} {2015})}\BibitemShut {NoStop}%
\bibitem [{\citenamefont {Osada}\ \emph {et~al.}(2018)\citenamefont {Osada},
  \citenamefont {Gloppe}, \citenamefont {Hisatomi}, \citenamefont {Noguchi},
  \citenamefont {Yamazaki}, \citenamefont {Nomura}, \citenamefont {Nakamura},\
  and\ \citenamefont {Usami}}]{Osada2018}%
  \BibitemOpen
  \bibfield  {author} {\bibinfo {author} {\bibfnamefont {A.}~\bibnamefont
  {Osada}}, \bibinfo {author} {\bibfnamefont {A.}~\bibnamefont {Gloppe}},
  \bibinfo {author} {\bibfnamefont {R.}~\bibnamefont {Hisatomi}}, \bibinfo
  {author} {\bibfnamefont {A.}~\bibnamefont {Noguchi}}, \bibinfo {author}
  {\bibfnamefont {R.}~\bibnamefont {Yamazaki}}, \bibinfo {author}
  {\bibfnamefont {M.}~\bibnamefont {Nomura}}, \bibinfo {author} {\bibfnamefont
  {Y.}~\bibnamefont {Nakamura}},\ and\ \bibinfo {author} {\bibfnamefont
  {K.}~\bibnamefont {Usami}},\ }\bibfield  {title} {\bibinfo {title} {Brillouin
  light scattering by magnetic quasivortices in cavity optomagnonics},\ }\href
  {https://doi.org/10.1103/PhysRevLett.120.133602} {\bibfield  {journal}
  {\bibinfo  {journal} {Phys. Rev. Lett.}\ }\textbf {\bibinfo {volume} {120}},\
  \bibinfo {pages} {133602} (\bibinfo {year} {2018})}\BibitemShut {NoStop}%
\bibitem [{\citenamefont {Haigh}\ \emph {et~al.}(2016)\citenamefont {Haigh},
  \citenamefont {Nunnenkamp}, \citenamefont {Ramsay},\ and\ \citenamefont
  {Ferguson}}]{Haigh2016}%
  \BibitemOpen
  \bibfield  {author} {\bibinfo {author} {\bibfnamefont {J.~A.}\ \bibnamefont
  {Haigh}}, \bibinfo {author} {\bibfnamefont {A.}~\bibnamefont {Nunnenkamp}},
  \bibinfo {author} {\bibfnamefont {A.~J.}\ \bibnamefont {Ramsay}},\ and\
  \bibinfo {author} {\bibfnamefont {A.~J.}\ \bibnamefont {Ferguson}},\
  }\bibfield  {title} {\bibinfo {title} {Triple-resonant brillouin light
  scattering in magneto-optical cavities},\ }\href
  {https://doi.org/10.1103/PhysRevLett.117.133602} {\bibfield  {journal}
  {\bibinfo  {journal} {Phys. Rev. Lett.}\ }\textbf {\bibinfo {volume} {117}},\
  \bibinfo {pages} {133602} (\bibinfo {year} {2016})}\BibitemShut {NoStop}%
\bibitem [{\citenamefont {Tabuchi}\ \emph {et~al.}(2015)\citenamefont
  {Tabuchi}, \citenamefont {Ishino}, \citenamefont {Noguchi}, \citenamefont
  {Ishikawa}, \citenamefont {Yamazaki}, \citenamefont {Usami},\ and\
  \citenamefont {Nakamura}}]{tabuchi2015coherent}%
  \BibitemOpen
  \bibfield  {author} {\bibinfo {author} {\bibfnamefont {Y.}~\bibnamefont
  {Tabuchi}}, \bibinfo {author} {\bibfnamefont {S.}~\bibnamefont {Ishino}},
  \bibinfo {author} {\bibfnamefont {A.}~\bibnamefont {Noguchi}}, \bibinfo
  {author} {\bibfnamefont {T.}~\bibnamefont {Ishikawa}}, \bibinfo {author}
  {\bibfnamefont {R.}~\bibnamefont {Yamazaki}}, \bibinfo {author}
  {\bibfnamefont {K.}~\bibnamefont {Usami}},\ and\ \bibinfo {author}
  {\bibfnamefont {Y.}~\bibnamefont {Nakamura}},\ }\bibfield  {title} {\bibinfo
  {title} {Coherent coupling between a ferromagnetic magnon and a
  superconducting qubit},\ }\href
  {https://doi.org/https://doi.org/10.1126/science.aaa3693} {\bibfield
  {journal} {\bibinfo  {journal} {Science}\ }\textbf {\bibinfo {volume}
  {349}},\ \bibinfo {pages} {405} (\bibinfo {year} {2015})}\BibitemShut
  {NoStop}%
\bibitem [{\citenamefont {Wolski}\ \emph {et~al.}(2020)\citenamefont {Wolski},
  \citenamefont {Lachance-Quirion}, \citenamefont {Tabuchi}, \citenamefont
  {Kono}, \citenamefont {Noguchi}, \citenamefont {Usami},\ and\ \citenamefont
  {Nakamura}}]{Wolski2020}%
  \BibitemOpen
  \bibfield  {author} {\bibinfo {author} {\bibfnamefont {S.~P.}\ \bibnamefont
  {Wolski}}, \bibinfo {author} {\bibfnamefont {D.}~\bibnamefont
  {Lachance-Quirion}}, \bibinfo {author} {\bibfnamefont {Y.}~\bibnamefont
  {Tabuchi}}, \bibinfo {author} {\bibfnamefont {S.}~\bibnamefont {Kono}},
  \bibinfo {author} {\bibfnamefont {A.}~\bibnamefont {Noguchi}}, \bibinfo
  {author} {\bibfnamefont {K.}~\bibnamefont {Usami}},\ and\ \bibinfo {author}
  {\bibfnamefont {Y.}~\bibnamefont {Nakamura}},\ }\bibfield  {title} {\bibinfo
  {title} {Dissipation-based quantum sensing of magnons with a superconducting
  qubit},\ }\href {https://doi.org/10.1103/PhysRevLett.125.117701} {\bibfield
  {journal} {\bibinfo  {journal} {Phys. Rev. Lett.}\ }\textbf {\bibinfo
  {volume} {125}},\ \bibinfo {pages} {117701} (\bibinfo {year}
  {2020})}\BibitemShut {NoStop}%
\bibitem [{\citenamefont {Li}\ \emph {et~al.}(2022)\citenamefont {Li},
  \citenamefont {Yefremenko}, \citenamefont {Lisovenko}, \citenamefont
  {Trevillian}, \citenamefont {Polakovic}, \citenamefont {Cecil}, \citenamefont
  {Barry}, \citenamefont {Pearson}, \citenamefont {Divan}, \citenamefont
  {Tyberkevych}, \citenamefont {Chang}, \citenamefont {Welp}, \citenamefont
  {Kwok},\ and\ \citenamefont {Novosad}}]{Li2022}%
  \BibitemOpen
  \bibfield  {author} {\bibinfo {author} {\bibfnamefont {Y.}~\bibnamefont
  {Li}}, \bibinfo {author} {\bibfnamefont {V.~G.}\ \bibnamefont {Yefremenko}},
  \bibinfo {author} {\bibfnamefont {M.}~\bibnamefont {Lisovenko}}, \bibinfo
  {author} {\bibfnamefont {C.}~\bibnamefont {Trevillian}}, \bibinfo {author}
  {\bibfnamefont {T.}~\bibnamefont {Polakovic}}, \bibinfo {author}
  {\bibfnamefont {T.~W.}\ \bibnamefont {Cecil}}, \bibinfo {author}
  {\bibfnamefont {P.~S.}\ \bibnamefont {Barry}}, \bibinfo {author}
  {\bibfnamefont {J.}~\bibnamefont {Pearson}}, \bibinfo {author} {\bibfnamefont
  {R.}~\bibnamefont {Divan}}, \bibinfo {author} {\bibfnamefont
  {V.}~\bibnamefont {Tyberkevych}}, \bibinfo {author} {\bibfnamefont {C.~L.}\
  \bibnamefont {Chang}}, \bibinfo {author} {\bibfnamefont {U.}~\bibnamefont
  {Welp}}, \bibinfo {author} {\bibfnamefont {W.-K.}\ \bibnamefont {Kwok}},\
  and\ \bibinfo {author} {\bibfnamefont {V.}~\bibnamefont {Novosad}},\
  }\bibfield  {title} {\bibinfo {title} {Coherent coupling of two remote
  magnonic resonators mediated by superconducting circuits},\ }\href
  {https://doi.org/10.1103/PhysRevLett.128.047701} {\bibfield  {journal}
  {\bibinfo  {journal} {Phys. Rev. Lett.}\ }\textbf {\bibinfo {volume} {128}},\
  \bibinfo {pages} {047701} (\bibinfo {year} {2022})}\BibitemShut {NoStop}%
\bibitem [{\citenamefont {Sun}\ \emph {et~al.}(2021)\citenamefont {Sun},
  \citenamefont {Zheng}, \citenamefont {Xiao}, \citenamefont {Gong},
  \citenamefont {He},\ and\ \citenamefont {Xia}}]{Sun2021}%
  \BibitemOpen
  \bibfield  {author} {\bibinfo {author} {\bibfnamefont {F.-X.}\ \bibnamefont
  {Sun}}, \bibinfo {author} {\bibfnamefont {S.-S.}\ \bibnamefont {Zheng}},
  \bibinfo {author} {\bibfnamefont {Y.}~\bibnamefont {Xiao}}, \bibinfo {author}
  {\bibfnamefont {Q.}~\bibnamefont {Gong}}, \bibinfo {author} {\bibfnamefont
  {Q.}~\bibnamefont {He}},\ and\ \bibinfo {author} {\bibfnamefont
  {K.}~\bibnamefont {Xia}},\ }\bibfield  {title} {\bibinfo {title} {Remote
  generation of magnon schr\"odinger cat state via magnon-photon
  entanglement},\ }\href {https://doi.org/10.1103/PhysRevLett.127.087203}
  {\bibfield  {journal} {\bibinfo  {journal} {Phys. Rev. Lett.}\ }\textbf
  {\bibinfo {volume} {127}},\ \bibinfo {pages} {087203} (\bibinfo {year}
  {2021})}\BibitemShut {NoStop}%
\bibitem [{\citenamefont {Huang}\ \emph {et~al.}(2023)\citenamefont {Huang},
  \citenamefont {Deng},\ and\ \citenamefont {Chen}}]{Huang2023}%
  \BibitemOpen
  \bibfield  {author} {\bibinfo {author} {\bibfnamefont {S.}~\bibnamefont
  {Huang}}, \bibinfo {author} {\bibfnamefont {L.}~\bibnamefont {Deng}},\ and\
  \bibinfo {author} {\bibfnamefont {A.}~\bibnamefont {Chen}},\ }\bibfield
  {title} {\bibinfo {title} {Squeezed states of magnons and phonons in a cavity
  magnomechanical system with a microwave parametric amplifier},\ }\href
  {https://doi.org/https://doi.org/10.1002/andp.202300095} {\bibfield
  {journal} {\bibinfo  {journal} {Ann. Phys. (Berlin)}\ }\textbf {\bibinfo
  {volume} {535}},\ \bibinfo {pages} {2300095} (\bibinfo {year}
  {2023})}\BibitemShut {NoStop}%
\bibitem [{\citenamefont {Potts}\ \emph {et~al.}(2021)\citenamefont {Potts},
  \citenamefont {Varga}, \citenamefont {Bittencourt}, \citenamefont
  {Kusminskiy},\ and\ \citenamefont {Davis}}]{potts2021dynamical}%
  \BibitemOpen
  \bibfield  {author} {\bibinfo {author} {\bibfnamefont {C.~A.}\ \bibnamefont
  {Potts}}, \bibinfo {author} {\bibfnamefont {E.}~\bibnamefont {Varga}},
  \bibinfo {author} {\bibfnamefont {V.~A.}\ \bibnamefont {Bittencourt}},
  \bibinfo {author} {\bibfnamefont {S.~V.}\ \bibnamefont {Kusminskiy}},\ and\
  \bibinfo {author} {\bibfnamefont {J.~P.}\ \bibnamefont {Davis}},\ }\bibfield
  {title} {\bibinfo {title} {Dynamical backaction magnomechanics},\ }\href
  {https://doi.org/https://doi.org/10.1103/PhysRevX.11.031053} {\bibfield
  {journal} {\bibinfo  {journal} {Phys. Rev. X}\ }\textbf {\bibinfo {volume}
  {11}},\ \bibinfo {pages} {031053} (\bibinfo {year} {2021})}\BibitemShut
  {NoStop}%
\bibitem [{\citenamefont {Li}\ \emph {et~al.}(2018)\citenamefont {Li},
  \citenamefont {Zhu},\ and\ \citenamefont {Agarwal}}]{li2018magnon}%
  \BibitemOpen
  \bibfield  {author} {\bibinfo {author} {\bibfnamefont {J.}~\bibnamefont
  {Li}}, \bibinfo {author} {\bibfnamefont {S.-Y.}\ \bibnamefont {Zhu}},\ and\
  \bibinfo {author} {\bibfnamefont {G.}~\bibnamefont {Agarwal}},\ }\bibfield
  {title} {\bibinfo {title} {Magnon-photon-phonon entanglement in cavity
  magnomechanics},\ }\href
  {https://doi.org/https://doi.org/10.1103/PhysRevLett.121.203601} {\bibfield
  {journal} {\bibinfo  {journal} {Phys. Rev. Lett.}\ }\textbf {\bibinfo
  {volume} {121}},\ \bibinfo {pages} {203601} (\bibinfo {year}
  {2018})}\BibitemShut {NoStop}%
\bibitem [{\citenamefont {Liu}\ \emph {et~al.}(2023)\citenamefont {Liu},
  \citenamefont {Liu}, \citenamefont {Tan},\ and\ \citenamefont
  {Liu}}]{Liu2023}%
  \BibitemOpen
  \bibfield  {author} {\bibinfo {author} {\bibfnamefont {Z.-Q.}\ \bibnamefont
  {Liu}}, \bibinfo {author} {\bibfnamefont {Y.}~\bibnamefont {Liu}}, \bibinfo
  {author} {\bibfnamefont {L.}~\bibnamefont {Tan}},\ and\ \bibinfo {author}
  {\bibfnamefont {W.-M.}\ \bibnamefont {Liu}},\ }\bibfield  {title} {\bibinfo
  {title} {Reservoir engineering strong magnomechanical entanglement via
  dual-mode cooling},\ }\href
  {https://doi.org/https://doi.org/10.1002/andp.202200660} {\bibfield
  {journal} {\bibinfo  {journal} {Ann. Phys. (Berlin)}\ }\textbf {\bibinfo
  {volume} {535}},\ \bibinfo {pages} {2200660} (\bibinfo {year}
  {2023})}\BibitemShut {NoStop}%
\bibitem [{\citenamefont {Asjad}\ \emph {et~al.}(2023)\citenamefont {Asjad},
  \citenamefont {Li}, \citenamefont {Zhu},\ and\ \citenamefont
  {You}}]{Asjad2022}%
  \BibitemOpen
  \bibfield  {author} {\bibinfo {author} {\bibfnamefont {M.}~\bibnamefont
  {Asjad}}, \bibinfo {author} {\bibfnamefont {J.}~\bibnamefont {Li}}, \bibinfo
  {author} {\bibfnamefont {S.-Y.}\ \bibnamefont {Zhu}},\ and\ \bibinfo {author}
  {\bibfnamefont {J.}~\bibnamefont {You}},\ }\bibfield  {title} {\bibinfo
  {title} {Magnon squeezing enhanced ground-state cooling in cavity
  magnomechanics},\ }\href
  {https://doi.org/https://doi.org/10.1016/j.fmre.2022.07.006} {\bibfield
  {journal} {\bibinfo  {journal} {Fundamental Research}\ }\textbf {\bibinfo
  {volume} {3}},\ \bibinfo {pages} {3} (\bibinfo {year} {2023})}\BibitemShut
  {NoStop}%
\bibitem [{\citenamefont {Tan}(2019)}]{Tan2019}%
  \BibitemOpen
  \bibfield  {author} {\bibinfo {author} {\bibfnamefont {H.}~\bibnamefont
  {Tan}},\ }\bibfield  {title} {\bibinfo {title} {Genuine photon-magnon-phonon
  einstein-podolsky-rosen steerable nonlocality in a continuously-monitored
  cavity magnomechanical system},\ }\href
  {https://doi.org/10.1103/PhysRevResearch.1.033161} {\bibfield  {journal}
  {\bibinfo  {journal} {Phys. Rev. Res.}\ }\textbf {\bibinfo {volume} {1}},\
  \bibinfo {pages} {033161} (\bibinfo {year} {2019})}\BibitemShut {NoStop}%
\bibitem [{\citenamefont {Li}\ \emph {et~al.}(2020)\citenamefont {Li},
  \citenamefont {Zhang}, \citenamefont {Tyberkevych}, \citenamefont {Kwok},
  \citenamefont {Hoffmann},\ and\ \citenamefont {Novosad}}]{Li2020}%
  \BibitemOpen
  \bibfield  {author} {\bibinfo {author} {\bibfnamefont {Y.}~\bibnamefont
  {Li}}, \bibinfo {author} {\bibfnamefont {W.}~\bibnamefont {Zhang}}, \bibinfo
  {author} {\bibfnamefont {V.}~\bibnamefont {Tyberkevych}}, \bibinfo {author}
  {\bibfnamefont {W.-K.}\ \bibnamefont {Kwok}}, \bibinfo {author}
  {\bibfnamefont {A.}~\bibnamefont {Hoffmann}},\ and\ \bibinfo {author}
  {\bibfnamefont {V.}~\bibnamefont {Novosad}},\ }\bibfield  {title} {\bibinfo
  {title} {{Hybrid magnonics: Physics, circuits, and applications for coherent
  information processing}},\ }\bibfield  {journal} {\bibinfo  {journal}
  {Journal of Applied Physics}\ }\textbf {\bibinfo {volume} {128}},\ \href
  {https://doi.org/10.1063/5.0020277} {10.1063/5.0020277} (\bibinfo {year}
  {2020})\BibitemShut {NoStop}%
\bibitem [{\citenamefont {Li}\ \emph {et~al.}(2021)\citenamefont {Li},
  \citenamefont {Wang}, \citenamefont {Wu}, \citenamefont {Zhu},\ and\
  \citenamefont {You}}]{Li2021}%
  \BibitemOpen
  \bibfield  {author} {\bibinfo {author} {\bibfnamefont {J.}~\bibnamefont
  {Li}}, \bibinfo {author} {\bibfnamefont {Y.-P.}\ \bibnamefont {Wang}},
  \bibinfo {author} {\bibfnamefont {W.-J.}\ \bibnamefont {Wu}}, \bibinfo
  {author} {\bibfnamefont {S.-Y.}\ \bibnamefont {Zhu}},\ and\ \bibinfo {author}
  {\bibfnamefont {J.}~\bibnamefont {You}},\ }\bibfield  {title} {\bibinfo
  {title} {Quantum network with magnonic and mechanical nodes},\ }\href
  {https://doi.org/10.1103/PRXQuantum.2.040344} {\bibfield  {journal} {\bibinfo
   {journal} {PRX Quantum}\ }\textbf {\bibinfo {volume} {2}},\ \bibinfo {pages}
  {040344} (\bibinfo {year} {2021})}\BibitemShut {NoStop}%
\bibitem [{\citenamefont {Yuan}\ \emph {et~al.}(2022)\citenamefont {Yuan},
  \citenamefont {Cao}, \citenamefont {Kamra}, \citenamefont {Duine},\ and\
  \citenamefont {Yan}}]{Yuan2022}%
  \BibitemOpen
  \bibfield  {author} {\bibinfo {author} {\bibfnamefont {H.}~\bibnamefont
  {Yuan}}, \bibinfo {author} {\bibfnamefont {Y.}~\bibnamefont {Cao}}, \bibinfo
  {author} {\bibfnamefont {A.}~\bibnamefont {Kamra}}, \bibinfo {author}
  {\bibfnamefont {R.~A.}\ \bibnamefont {Duine}},\ and\ \bibinfo {author}
  {\bibfnamefont {P.}~\bibnamefont {Yan}},\ }\bibfield  {title} {\bibinfo
  {title} {Quantum magnonics: When magnon spintronics meets quantum information
  science},\ }\href
  {https://doi.org/https://doi.org/10.1016/j.physrep.2022.03.002} {\bibfield
  {journal} {\bibinfo  {journal} {Phys. Rep.}\ }\textbf {\bibinfo {volume}
  {965}},\ \bibinfo {pages} {1} (\bibinfo {year} {2022})}\BibitemShut {NoStop}%
\bibitem [{\citenamefont {Goold}\ \emph {et~al.}(2016)\citenamefont {Goold},
  \citenamefont {Huber}, \citenamefont {Riera}, \citenamefont {del Rio},\ and\
  \citenamefont {Skrzypczyk}}]{Goold2016}%
  \BibitemOpen
  \bibfield  {author} {\bibinfo {author} {\bibfnamefont {J.}~\bibnamefont
  {Goold}}, \bibinfo {author} {\bibfnamefont {M.}~\bibnamefont {Huber}},
  \bibinfo {author} {\bibfnamefont {A.}~\bibnamefont {Riera}}, \bibinfo
  {author} {\bibfnamefont {L.}~\bibnamefont {del Rio}},\ and\ \bibinfo {author}
  {\bibfnamefont {P.}~\bibnamefont {Skrzypczyk}},\ }\bibfield  {title}
  {\bibinfo {title} {The role of quantum information in thermodynamics---a
  topical review},\ }\href {https://doi.org/10.1088/1751-8113/49/14/143001}
  {\bibfield  {journal} {\bibinfo  {journal} {J. Phys. A Math. Theor.}\
  }\textbf {\bibinfo {volume} {49}},\ \bibinfo {pages} {143001} (\bibinfo
  {year} {2016})}\BibitemShut {NoStop}%
\bibitem [{\citenamefont {Zhang}\ \emph
  {et~al.}(2014{\natexlab{a}})\citenamefont {Zhang}, \citenamefont {Bariani},\
  and\ \citenamefont {Meystre}}]{zhang2014quantum}%
  \BibitemOpen
  \bibfield  {author} {\bibinfo {author} {\bibfnamefont {K.}~\bibnamefont
  {Zhang}}, \bibinfo {author} {\bibfnamefont {F.}~\bibnamefont {Bariani}},\
  and\ \bibinfo {author} {\bibfnamefont {P.}~\bibnamefont {Meystre}},\
  }\bibfield  {title} {\bibinfo {title} {Quantum optomechanical heat engine},\
  }\href {https://doi.org/https://doi.org/10.1103/PhysRevLett.112.150602}
  {\bibfield  {journal} {\bibinfo  {journal} {Phys. Rev. Lett.}\ }\textbf
  {\bibinfo {volume} {112}},\ \bibinfo {pages} {150602} (\bibinfo {year}
  {2014}{\natexlab{a}})}\BibitemShut {NoStop}%
\bibitem [{\citenamefont {Zhang}\ \emph
  {et~al.}(2014{\natexlab{b}})\citenamefont {Zhang}, \citenamefont {Bariani},\
  and\ \citenamefont {Meystre}}]{zhang2014theory}%
  \BibitemOpen
  \bibfield  {author} {\bibinfo {author} {\bibfnamefont {K.}~\bibnamefont
  {Zhang}}, \bibinfo {author} {\bibfnamefont {F.}~\bibnamefont {Bariani}},\
  and\ \bibinfo {author} {\bibfnamefont {P.}~\bibnamefont {Meystre}},\
  }\bibfield  {title} {\bibinfo {title} {Theory of an optomechanical quantum
  heat engine},\ }\href
  {https://doi.org/https://doi.org/10.1103/PhysRevA.90.023819} {\bibfield
  {journal} {\bibinfo  {journal} {Phys. Rev. A}\ }\textbf {\bibinfo {volume}
  {90}},\ \bibinfo {pages} {023819} (\bibinfo {year}
  {2014}{\natexlab{b}})}\BibitemShut {NoStop}%
\bibitem [{\citenamefont {Zhang}\ and\ \citenamefont
  {Zhang}(2017)}]{Zhang2017}%
  \BibitemOpen
  \bibfield  {author} {\bibinfo {author} {\bibfnamefont {K.}~\bibnamefont
  {Zhang}}\ and\ \bibinfo {author} {\bibfnamefont {W.}~\bibnamefont {Zhang}},\
  }\bibfield  {title} {\bibinfo {title} {Quantum optomechanical straight-twin
  engine},\ }\href {https://doi.org/10.1103/PhysRevA.95.053870} {\bibfield
  {journal} {\bibinfo  {journal} {Phys. Rev. A}\ }\textbf {\bibinfo {volume}
  {95}},\ \bibinfo {pages} {053870} (\bibinfo {year} {2017})}\BibitemShut
  {NoStop}%
\bibitem [{\citenamefont {Dechant}\ \emph {et~al.}(2015)\citenamefont
  {Dechant}, \citenamefont {Kiesel},\ and\ \citenamefont {Lutz}}]{Dechant2015}%
  \BibitemOpen
  \bibfield  {author} {\bibinfo {author} {\bibfnamefont {A.}~\bibnamefont
  {Dechant}}, \bibinfo {author} {\bibfnamefont {N.}~\bibnamefont {Kiesel}},\
  and\ \bibinfo {author} {\bibfnamefont {E.}~\bibnamefont {Lutz}},\ }\bibfield
  {title} {\bibinfo {title} {All-optical nanomechanical heat engine},\ }\href
  {https://doi.org/10.1103/PhysRevLett.114.183602} {\bibfield  {journal}
  {\bibinfo  {journal} {Phys. Rev. Lett.}\ }\textbf {\bibinfo {volume} {114}},\
  \bibinfo {pages} {183602} (\bibinfo {year} {2015})}\BibitemShut {NoStop}%
\bibitem [{\citenamefont {Serafini}\ \emph {et~al.}(2020)\citenamefont
  {Serafini}, \citenamefont {Zippilli},\ and\ \citenamefont
  {Marzoli}}]{Serafini2020}%
  \BibitemOpen
  \bibfield  {author} {\bibinfo {author} {\bibfnamefont {G.}~\bibnamefont
  {Serafini}}, \bibinfo {author} {\bibfnamefont {S.}~\bibnamefont {Zippilli}},\
  and\ \bibinfo {author} {\bibfnamefont {I.}~\bibnamefont {Marzoli}},\
  }\bibfield  {title} {\bibinfo {title} {Optomechanical stirling heat engine
  driven by feedback-controlled light},\ }\href
  {https://doi.org/10.1103/PhysRevA.102.053502} {\bibfield  {journal} {\bibinfo
   {journal} {Phys. Rev. A}\ }\textbf {\bibinfo {volume} {102}},\ \bibinfo
  {pages} {053502} (\bibinfo {year} {2020})}\BibitemShut {NoStop}%
\bibitem [{\citenamefont {Bathaee}\ and\ \citenamefont
  {Bahrampour}(2016)}]{Bathaee2016}%
  \BibitemOpen
  \bibfield  {author} {\bibinfo {author} {\bibfnamefont {M.}~\bibnamefont
  {Bathaee}}\ and\ \bibinfo {author} {\bibfnamefont {A.~R.}\ \bibnamefont
  {Bahrampour}},\ }\bibfield  {title} {\bibinfo {title} {Optimal control of the
  power adiabatic stroke of an optomechanical heat engine},\ }\href
  {https://doi.org/10.1103/PhysRevE.94.022141} {\bibfield  {journal} {\bibinfo
  {journal} {Phys. Rev. E}\ }\textbf {\bibinfo {volume} {94}},\ \bibinfo
  {pages} {022141} (\bibinfo {year} {2016})}\BibitemShut {NoStop}%
\bibitem [{\citenamefont {Naseem}\ and\ \citenamefont {\"{O}zg\"{u}r
  E.~M\"{u}stecaplio\u{g}lu}(2019)}]{Naseem2019}%
  \BibitemOpen
  \bibfield  {author} {\bibinfo {author} {\bibfnamefont {M.~T.}\ \bibnamefont
  {Naseem}}\ and\ \bibinfo {author} {\bibnamefont {\"{O}zg\"{u}r
  E.~M\"{u}stecaplio\u{g}lu}},\ }\bibfield  {title} {\bibinfo {title} {Quantum
  heat engine with a quadratically coupled optomechanical system},\ }\href
  {https://doi.org/10.1364/JOSAB.36.003000} {\bibfield  {journal} {\bibinfo
  {journal} {J. Opt. Soc. Am. B}\ }\textbf {\bibinfo {volume} {36}},\ \bibinfo
  {pages} {3000} (\bibinfo {year} {2019})}\BibitemShut {NoStop}%
\bibitem [{\citenamefont {Tolman}\ and\ \citenamefont
  {Fine}(1948)}]{tolman1948irreversible}%
  \BibitemOpen
  \bibfield  {author} {\bibinfo {author} {\bibfnamefont {R.~C.}\ \bibnamefont
  {Tolman}}\ and\ \bibinfo {author} {\bibfnamefont {P.~C.}\ \bibnamefont
  {Fine}},\ }\bibfield  {title} {\bibinfo {title} {On the irreversible
  production of entropy},\ }\href
  {https://doi.org/https://doi.org/10.1103/RevModPhys.20.51} {\bibfield
  {journal} {\bibinfo  {journal} {Rev. Mod. Phys.}\ }\textbf {\bibinfo {volume}
  {20}},\ \bibinfo {pages} {51} (\bibinfo {year} {1948})}\BibitemShut {NoStop}%
\bibitem [{\citenamefont {Sethna}(2021)}]{sethna2021statistical}%
  \BibitemOpen
  \bibfield  {author} {\bibinfo {author} {\bibfnamefont {J.~P.}\ \bibnamefont
  {Sethna}},\ }\href {https://academic.oup.com/book/44296} {\emph {\bibinfo
  {title} {Statistical mechanics: entropy, order parameters, and
  complexity}}},\ Vol.~\bibinfo {volume} {14}\ (\bibinfo  {publisher} {Oxford
  University Press, USA},\ \bibinfo {year} {2021})\BibitemShut {NoStop}%
\bibitem [{\citenamefont {Landi}\ and\ \citenamefont
  {Paternostro}(2021)}]{landi2021irreversible}%
  \BibitemOpen
  \bibfield  {author} {\bibinfo {author} {\bibfnamefont {G.~T.}\ \bibnamefont
  {Landi}}\ and\ \bibinfo {author} {\bibfnamefont {M.}~\bibnamefont
  {Paternostro}},\ }\bibfield  {title} {\bibinfo {title} {Irreversible entropy
  production: From classical to quantum},\ }\href
  {https://doi.org/https://doi.org/10.1103/RevModPhys.93.035008} {\bibfield
  {journal} {\bibinfo  {journal} {Rev. Mod. Phys.}\ }\textbf {\bibinfo {volume}
  {93}},\ \bibinfo {pages} {035008} (\bibinfo {year} {2021})}\BibitemShut
  {NoStop}%
\bibitem [{\citenamefont {Brunelli}\ and\ \citenamefont
  {Paternostro}(2016)}]{brunelli2016irreversibility}%
  \BibitemOpen
  \bibfield  {author} {\bibinfo {author} {\bibfnamefont {M.}~\bibnamefont
  {Brunelli}}\ and\ \bibinfo {author} {\bibfnamefont {M.}~\bibnamefont
  {Paternostro}},\ }\bibfield  {title} {\bibinfo {title} {Irreversibility and
  correlations in coupled quantum oscillators},\ }\bibfield  {journal}
  {\bibinfo  {journal} {arXiv preprint arXiv:1610.01172}\ }\href
  {https://doi.org/https://doi.org/10.48550/arXiv.1610.01172}
  {https://doi.org/10.48550/arXiv.1610.01172} (\bibinfo {year}
  {2016})\BibitemShut {NoStop}%
\bibitem [{\citenamefont {Santos}\ \emph {et~al.}(2017)\citenamefont {Santos},
  \citenamefont {Landi},\ and\ \citenamefont {Paternostro}}]{santos2017wigner}%
  \BibitemOpen
  \bibfield  {author} {\bibinfo {author} {\bibfnamefont {J.~P.}\ \bibnamefont
  {Santos}}, \bibinfo {author} {\bibfnamefont {G.~T.}\ \bibnamefont {Landi}},\
  and\ \bibinfo {author} {\bibfnamefont {M.}~\bibnamefont {Paternostro}},\
  }\bibfield  {title} {\bibinfo {title} {Wigner entropy production rate},\
  }\href {https://doi.org/https://doi.org/10.1103/PhysRevLett.118.220601}
  {\bibfield  {journal} {\bibinfo  {journal} {Phys. Rev. Lett.}\ }\textbf
  {\bibinfo {volume} {118}},\ \bibinfo {pages} {220601} (\bibinfo {year}
  {2017})}\BibitemShut {NoStop}%
\bibitem [{\citenamefont {Santos}\ \emph {et~al.}(2018)\citenamefont {Santos},
  \citenamefont {C{\'e}leri}, \citenamefont {Brito}, \citenamefont {Landi},\
  and\ \citenamefont {Paternostro}}]{santos2018spin}%
  \BibitemOpen
  \bibfield  {author} {\bibinfo {author} {\bibfnamefont {J.~P.}\ \bibnamefont
  {Santos}}, \bibinfo {author} {\bibfnamefont {L.~C.}\ \bibnamefont
  {C{\'e}leri}}, \bibinfo {author} {\bibfnamefont {F.}~\bibnamefont {Brito}},
  \bibinfo {author} {\bibfnamefont {G.~T.}\ \bibnamefont {Landi}},\ and\
  \bibinfo {author} {\bibfnamefont {M.}~\bibnamefont {Paternostro}},\
  }\bibfield  {title} {\bibinfo {title} {Spin-phase-space-entropy production},\
  }\href {https://doi.org/https://doi.org/10.1103/PhysRevA.97.052123}
  {\bibfield  {journal} {\bibinfo  {journal} {Phys. Rev. A}\ }\textbf {\bibinfo
  {volume} {97}},\ \bibinfo {pages} {052123} (\bibinfo {year}
  {2018})}\BibitemShut {NoStop}%
\bibitem [{\citenamefont {Brunelli}\ \emph {et~al.}(2018)\citenamefont
  {Brunelli}, \citenamefont {Fusco}, \citenamefont {Landig}, \citenamefont
  {Wieczorek}, \citenamefont {Hoelscher-Obermaier}, \citenamefont {Landi},
  \citenamefont {Semi{\~a}o}, \citenamefont {Ferraro}, \citenamefont {Kiesel},
  \citenamefont {Donner} \emph {et~al.}}]{brunelli2018experimental}%
  \BibitemOpen
  \bibfield  {author} {\bibinfo {author} {\bibfnamefont {M.}~\bibnamefont
  {Brunelli}}, \bibinfo {author} {\bibfnamefont {L.}~\bibnamefont {Fusco}},
  \bibinfo {author} {\bibfnamefont {R.}~\bibnamefont {Landig}}, \bibinfo
  {author} {\bibfnamefont {W.}~\bibnamefont {Wieczorek}}, \bibinfo {author}
  {\bibfnamefont {J.}~\bibnamefont {Hoelscher-Obermaier}}, \bibinfo {author}
  {\bibfnamefont {G.}~\bibnamefont {Landi}}, \bibinfo {author} {\bibfnamefont
  {F.}~\bibnamefont {Semi{\~a}o}}, \bibinfo {author} {\bibfnamefont
  {A.}~\bibnamefont {Ferraro}}, \bibinfo {author} {\bibfnamefont
  {N.}~\bibnamefont {Kiesel}}, \bibinfo {author} {\bibfnamefont
  {T.}~\bibnamefont {Donner}}, \emph {et~al.},\ }\bibfield  {title} {\bibinfo
  {title} {Experimental determination of irreversible entropy production in
  out-of-equilibrium mesoscopic quantum systems},\ }\href
  {https://doi.org/https://doi.org/10.1103/PhysRevLett.121.160604} {\bibfield
  {journal} {\bibinfo  {journal} {Phys. Rev. Lett.}\ }\textbf {\bibinfo
  {volume} {121}},\ \bibinfo {pages} {160604} (\bibinfo {year}
  {2018})}\BibitemShut {NoStop}%
\bibitem [{\citenamefont {Shahidani}\ and\ \citenamefont
  {Rafiee}(2022)}]{shahidani2022irreversible}%
  \BibitemOpen
  \bibfield  {author} {\bibinfo {author} {\bibfnamefont {S.}~\bibnamefont
  {Shahidani}}\ and\ \bibinfo {author} {\bibfnamefont {M.}~\bibnamefont
  {Rafiee}},\ }\bibfield  {title} {\bibinfo {title} {Irreversible entropy
  production rate in a parametrically driven-dissipative system: The role of
  self-correlation between noncommuting observables},\ }\href
  {https://doi.org/https://doi.org/10.1103/PhysRevA.105.052430} {\bibfield
  {journal} {\bibinfo  {journal} {Phys. Rev. A}\ }\textbf {\bibinfo {volume}
  {105}},\ \bibinfo {pages} {052430} (\bibinfo {year} {2022})}\BibitemShut
  {NoStop}%
\bibitem [{\citenamefont {Abah}\ \emph {et~al.}()\citenamefont {Abah},
  \citenamefont {Edet}, \citenamefont {Ali}, \citenamefont {Teklu},\ and\
  \citenamefont {Asjad}}]{abah2023}%
  \BibitemOpen
  \bibfield  {author} {\bibinfo {author} {\bibfnamefont {O.}~\bibnamefont
  {Abah}}, \bibinfo {author} {\bibfnamefont {C.~O.}\ \bibnamefont {Edet}},
  \bibinfo {author} {\bibfnamefont {N.}~\bibnamefont {Ali}}, \bibinfo {author}
  {\bibfnamefont {B.}~\bibnamefont {Teklu}},\ and\ \bibinfo {author}
  {\bibfnamefont {M.}~\bibnamefont {Asjad}},\ }\bibfield  {title} {\bibinfo
  {title} {Irreversibility in an optical parametric driven optomechanical
  system},\ }\href {https://doi.org/https://doi.org/10.1002/andp.202300400}
  {\bibfield  {journal} {\bibinfo  {journal} {Annalen der Physik}\ }\textbf
  {\bibinfo {volume} {n/a}},\ \bibinfo {pages} {2300400}},\ \Eprint
  {https://arxiv.org/abs/https://onlinelibrary.wiley.com/doi/pdf/10.1002/andp.202300400}
  {https://onlinelibrary.wiley.com/doi/pdf/10.1002/andp.202300400} \BibitemShut
  {NoStop}%
\bibitem [{\citenamefont {Wang}\ \emph {et~al.}(2016)\citenamefont {Wang},
  \citenamefont {Zhang}, \citenamefont {Zhang}, \citenamefont {Luo},
  \citenamefont {Xiong}, \citenamefont {Wang}, \citenamefont {Li},
  \citenamefont {Hu},\ and\ \citenamefont {You}}]{wang2016magnon}%
  \BibitemOpen
  \bibfield  {author} {\bibinfo {author} {\bibfnamefont {Y.-P.}\ \bibnamefont
  {Wang}}, \bibinfo {author} {\bibfnamefont {G.-Q.}\ \bibnamefont {Zhang}},
  \bibinfo {author} {\bibfnamefont {D.}~\bibnamefont {Zhang}}, \bibinfo
  {author} {\bibfnamefont {X.-Q.}\ \bibnamefont {Luo}}, \bibinfo {author}
  {\bibfnamefont {W.}~\bibnamefont {Xiong}}, \bibinfo {author} {\bibfnamefont
  {S.-P.}\ \bibnamefont {Wang}}, \bibinfo {author} {\bibfnamefont {T.-F.}\
  \bibnamefont {Li}}, \bibinfo {author} {\bibfnamefont {C.-M.}\ \bibnamefont
  {Hu}},\ and\ \bibinfo {author} {\bibfnamefont {J.}~\bibnamefont {You}},\
  }\bibfield  {title} {\bibinfo {title} {Magnon kerr effect in a strongly
  coupled cavity-magnon system},\ }\href
  {https://doi.org/https://doi.org/10.1103/PhysRevB.94.224410} {\bibfield
  {journal} {\bibinfo  {journal} {Phys. Rev. B}\ }\textbf {\bibinfo {volume}
  {94}},\ \bibinfo {pages} {224410} (\bibinfo {year} {2016})}\BibitemShut
  {NoStop}%
\bibitem [{\citenamefont {Ding}\ \emph {et~al.}(2020)\citenamefont {Ding},
  \citenamefont {Zheng},\ and\ \citenamefont {Li}}]{ding2020ground}%
  \BibitemOpen
  \bibfield  {author} {\bibinfo {author} {\bibfnamefont {M.-S.}\ \bibnamefont
  {Ding}}, \bibinfo {author} {\bibfnamefont {L.}~\bibnamefont {Zheng}},\ and\
  \bibinfo {author} {\bibfnamefont {C.}~\bibnamefont {Li}},\ }\bibfield
  {title} {\bibinfo {title} {Ground-state cooling of a magnomechanical
  resonator induced by magnetic damping},\ }\href
  {https://doi.org/https://doi.org/10.1364/JOSAB.380755} {\bibfield  {journal}
  {\bibinfo  {journal} {JOSA B}\ }\textbf {\bibinfo {volume} {37}},\ \bibinfo
  {pages} {627} (\bibinfo {year} {2020})}\BibitemShut {NoStop}%
\bibitem [{\citenamefont {Wang}\ \emph {et~al.}(2018)\citenamefont {Wang},
  \citenamefont {Zhang}, \citenamefont {Zhang}, \citenamefont {Li},
  \citenamefont {Hu},\ and\ \citenamefont {You}}]{wang2018bistability}%
  \BibitemOpen
  \bibfield  {author} {\bibinfo {author} {\bibfnamefont {Y.-P.}\ \bibnamefont
  {Wang}}, \bibinfo {author} {\bibfnamefont {G.-Q.}\ \bibnamefont {Zhang}},
  \bibinfo {author} {\bibfnamefont {D.}~\bibnamefont {Zhang}}, \bibinfo
  {author} {\bibfnamefont {T.-F.}\ \bibnamefont {Li}}, \bibinfo {author}
  {\bibfnamefont {C.-M.}\ \bibnamefont {Hu}},\ and\ \bibinfo {author}
  {\bibfnamefont {J.}~\bibnamefont {You}},\ }\bibfield  {title} {\bibinfo
  {title} {Bistability of cavity magnon polaritons},\ }\href
  {https://doi.org/https://doi.org/10.1103/PhysRevLett.120.057202} {\bibfield
  {journal} {\bibinfo  {journal} {Phys. Rev. Lett.}\ }\textbf {\bibinfo
  {volume} {120}},\ \bibinfo {pages} {057202} (\bibinfo {year}
  {2018})}\BibitemShut {NoStop}%
\bibitem [{\citenamefont {Kittel}(1948)}]{kittel1948theory}%
  \BibitemOpen
  \bibfield  {author} {\bibinfo {author} {\bibfnamefont {C.}~\bibnamefont
  {Kittel}},\ }\bibfield  {title} {\bibinfo {title} {On the theory of
  ferromagnetic resonance absorption},\ }\href
  {https://doi.org/https://doi.org/10.1103/PhysRev.73.155} {\bibfield
  {journal} {\bibinfo  {journal} {Phys. Rev.}\ }\textbf {\bibinfo {volume}
  {73}},\ \bibinfo {pages} {155} (\bibinfo {year} {1948})}\BibitemShut
  {NoStop}%
\bibitem [{\citenamefont {Gardiner}\ and\ \citenamefont
  {Zoller}(2004)}]{Gardiner2004}%
  \BibitemOpen
  \bibfield  {author} {\bibinfo {author} {\bibfnamefont {C.}~\bibnamefont
  {Gardiner}}\ and\ \bibinfo {author} {\bibfnamefont {P.}~\bibnamefont
  {Zoller}},\ }\href@noop {} {\emph {\bibinfo {title} {Quantum Noise}}}\
  (\bibinfo  {publisher} {Springer-Verlag},\ \bibinfo {address} {Berlin
  Heidelberg},\ \bibinfo {year} {2004})\BibitemShut {NoStop}%
\bibitem [{\citenamefont {DeJesus}\ and\ \citenamefont
  {Kaufman}(1987)}]{dejesus1987routh}%
  \BibitemOpen
  \bibfield  {author} {\bibinfo {author} {\bibfnamefont {E.~X.}\ \bibnamefont
  {DeJesus}}\ and\ \bibinfo {author} {\bibfnamefont {C.}~\bibnamefont
  {Kaufman}},\ }\bibfield  {title} {\bibinfo {title} {Routh-hurwitz criterion
  in the examination of eigenvalues of a system of nonlinear ordinary
  differential equations},\ }\href
  {https://doi.org/https://doi.org/10.1103/PhysRevA.35.5288} {\bibfield
  {journal} {\bibinfo  {journal} {Phys. Rev. A}\ }\textbf {\bibinfo {volume}
  {35}},\ \bibinfo {pages} {5288} (\bibinfo {year} {1987})}\BibitemShut
  {NoStop}%
\bibitem [{\citenamefont {Adesso}\ \emph {et~al.}(2012)\citenamefont {Adesso},
  \citenamefont {Girolami},\ and\ \citenamefont
  {Serafini}}]{adesso2012measuring}%
  \BibitemOpen
  \bibfield  {author} {\bibinfo {author} {\bibfnamefont {G.}~\bibnamefont
  {Adesso}}, \bibinfo {author} {\bibfnamefont {D.}~\bibnamefont {Girolami}},\
  and\ \bibinfo {author} {\bibfnamefont {A.}~\bibnamefont {Serafini}},\
  }\bibfield  {title} {\bibinfo {title} {Measuring gaussian quantum information
  and correlations using the r{\'e}nyi entropy of order 2},\ }\href
  {https://doi.org/https://doi.org/10.1103/PhysRevLett.109.190502} {\bibfield
  {journal} {\bibinfo  {journal} {Phys. Rev. Lett.}\ }\textbf {\bibinfo
  {volume} {109}},\ \bibinfo {pages} {190502} (\bibinfo {year}
  {2012})}\BibitemShut {NoStop}%
\end{thebibliography}
%

\end{document}